\documentclass[aip,jcp,reprint,amsmath,amssymb,floatfix,citeautoscript]{revtex4-1}
\usepackage{cancel}
\usepackage{amsmath}
\usepackage{physics}
\usepackage{graphicx}
\usepackage{amssymb}
\usepackage{amsthm}
\usepackage{bm}
\usepackage{dcolumn}
\usepackage{braket}
\usepackage{longtable}
\usepackage{ragged2e}
\usepackage{txfonts}
\usepackage[version=3]{mhchem} 
\usepackage[T1]{fontenc}       
\usepackage{pstricks}
\usepackage{algorithm}         
\usepackage{algpseudocode}     
\usepackage{siunitx}
\usepackage[colorlinks=true,allcolors=blue]{hyperref}
\usepackage[capitalise]{cleveref}

\let\tr\undefined
\newcommand*\tr[1]{\mathrm{#1}}
\newcommand*\mat[1]{\mathbf{#1}}

\newcommand*\rv[1]{{#1}}
\newcommand*\rvv[1]{{#1}}

\newcommand*\me{\mathrm{e}}
\newcommand*\md{\mathrm{d}}
\newcommand*\mi{\mathrm{i}}

\allowdisplaybreaks

\begin{document}

\title
{Fast periodic Gaussian density fitting by range separation}

\author{Hong-Zhou Ye}
\affiliation
{Department of Chemistry, Columbia University, New York, New York 10027, USA}
\author{Timothy C. Berkelbach}
\email{tim.berkelbach@gmail.com}
\affiliation
{Department of Chemistry, Columbia University, New York, New York 10027, USA}
\affiliation
{Center for Computational Quantum Physics, Flatiron Institute, New York, New York 10010, USA}

\begin{abstract}
    We present an efficient implementation of periodic Gaussian density fitting (GDF) using the Coulomb metric.
    The three-center integrals are divided into two parts by range-separating the Coulomb kernel, with the short-range part evaluated in real space and the long-range part in reciprocal space.
    With a few algorithmic optimizations, we show that this new method -- which we call range-separated GDF (RSGDF) -- scales sublinearly to linearly with the number of $k$-points for small to medium-sized $k$-point meshes that are commonly used in periodic calculations with electron correlation.
    Numerical results on a few three-dimensional solids show about $10$-fold speedups over the previously developed GDF with little precision loss.
    The error introduced by RSGDF is about $10^{-5}~E_{\tr{h}}$ in the converged Hartree-Fock energy
    with default auxiliary basis sets and can be systematically reduced by increasing the size of the auxiliary basis with little extra work.
\end{abstract}

\maketitle

\textit{Introduction}.
For one-electron basis sets in periodic electronic structure calculations, translational symmetry-adapted
atom-centered Gaussian functions\cite{Dovesi18WIRCMS,Sun18WIRCMS,Kuhne20JCP,Balasubramani20JCP} are an alternative to the historically prevalent plane waves \cite{Ihm79JPCSSP,Car85PRL,Martins88PRB,Pickett89CPR,Kresse96CMS}.
Using Gaussian basis functions provides a more compact representation of orbitals,
allows natural access to all-electron calculations without pseudopotentials, and facilitates the adaptation
of accurate quantum chemistry methods for solids. \cite{Maschio08PRB,Izmaylov08PCCP,Hirata09PRB,Maschio10JCP,Dovesi18WIRCMS,McClain17JCTC,Sun17JCP,Wang20JCTC,Buchholz18JCC}
The downside of atom-centered orbitals
is the introduction of four-index electron repulsion integrals (ERIs), with
$O(N_k^3 n^4_{\tr{AO}})$ storage and $O(N_k^2 n^4_{\tr{AO}})$ CPU costs for Hartree-Fock (HF) calculations, where $N_k$ is the number of $k$-points sampled in the Brillouin zone and
$n_{\tr{AO}}$ is the number of atomic orbitals in the unit cell. Moreover, the direct real-space evaluation of
ERIs requires an expensive triple lattice summation.
The Gaussian and plane wave (GPW) method \cite{VandeVondele05CPC} reduces the scaling of the
storage to $O(N_k^2n_{\tr{AO}}^2 N_{\tr{PW}})$
and the HF cost to $O(N_k^2n_{\tr{AO}}^2 N_{\tr{PW}} \ln N_{\tr{PW}})$
by evaluating the ERIs
entirely in reciprocal space using an auxiliary PW basis of size $N_{\tr{PW}}$. However,
doing so necessitates a pseudopotential and hence precludes all-electron
calculations. In addition, a large PW basis may be needed if the basis set
contains relatively compact orbitals.

Another way to reduce the cost of manipulating the ERIs is with Gaussian density fitting \cite{Whitten73JCP,Dunlap79JCP,Mintmire82PRA} (GDF). In GDF, the orbital pair densities used to evaluate the ERIs are
expanded in a second, auxiliary Gaussian basis of size $n_{\tr{aux}}$, from which the
four-center ERIs can be approximated using two-
and three-center integrals evaluated with some metric function \cite{Baerends73CP,Vahtras93CPL,Jung05PNAS,Reine08JCP}.
The number of the latter integrals scales as $O(N_k^2 n_{\tr{AO}}^2
n_{\tr{aux}})$, which is much lower than that of the ERIs if $n_{\tr{aux}}$ is
not too big. For molecules, highly optimized auxiliary basis sets \cite{Hill13IJQC} with
$n_{\tr{aux}} \approx 3 n_{\tr{AO}}$ have made GDF a great success in both
mean-field \cite{Sodt06JCP,Sodt08JCP,Manzer15JCP} and correlated calculations \cite{Werner03JCP,Eshuis10JCP,Werner11JCP,Riplinger13JCP,Gyorffy13JCP}.
We note that the GPW treatment of ERIs can also be understood as a PW density fitting
where $N_{\tr{PW}} \gg n_{\tr{AO}}$.

The application of GDF to periodic systems has been a relatively recent
effort. \cite{Varga05PRB,Varga06JCP,Maschio07PRB,Usvyat07PRB,Maschio08PRB,Pisani08JCC,Burow09JCP,Lazarski15JCTC,Luenser17JCTC,Grundei17JCTC,Wang20JCP} The main challenge is the high computational cost of
evaluating the three-center integrals in real space if the Coulomb metric
is used. There are two classes of
periodic GDF schemes.  The first class
exploits locality to limit the auxiliary
expansion based on the proximity to the target pair density. \cite{Maschio07PRB,Usvyat07PRB,Maschio08PRB,Pisani08JCC,Usvyat17inbook,Luenser17JCTC,Wang20JCP} The locality could
arise from the system itself \cite{Wang20JCP}, an explicit use of a local metric
other than the Coulomb operator \cite{Luenser17JCTC}, or \rv{the use of Poisson-type orbitals} \cite{Maschio07PRB,Usvyat07PRB,Maschio08PRB}.
The other class insists on a global, Coulomb metric-based GDF and accelerates
the integral evaluation by calculating the slowly convergent, long-range part
separately, e.g.,\ in reciprocal space using a PW basis \cite{Sun17JCP,Patterson20JCP} or in real space
using a multipole expansion \cite{Lazarski15JCTC,Lazarski16JCC,Becker19JCC,Burow09JCP,Grundei17JCTC}.
The global GDF with the Coulomb metric is generally considered more accurate but
less computationally efficient than the local one. \cite{Varga08IJQC,Varga11JMC,Merlot13JCC,Schmitz18CPL,Wirz17JCTC}

Here, we introduce an efficient implementation of a global,
Coulomb metric-based GDF for periodic systems. We use the error function to
range-separate the Coulomb metric integrals, evaluating the short-range part in
real space and the long-range part in reciprocal space, similar in spirit to
Refs.~\onlinecite{Shimazaki14JPSJ,Patterson20JCP,Sharma20arXiv}.
With a few algorithmic
developments, we show that the new scheme -- which we call range-separated Gaussian
density fitting (RSGDF) -- scales sublinearly to linearly with $N_k$ for small to
medium-sized $k$-point meshes that are commonly used in periodic
calculations with electron correlation \cite{Gruneis10JCP,Gruneis15JCP,Hummel16EPJB,McClain17JCTC,Wang20JCTC,MarySchmolzer20PRR}.
Numerical tests on three simple three-dimensional solids demonstrate that RSGDF
accelerates previous implementations of GDF by an order of magnitude with negligible
precision loss in the computed energies.
We also show that the accuracy of the HF \cite{Pisani80IJQC,Dovesi00PSSB} energy
computed using RSGDF can be systematically improved with little extra
computational effort by increasing the size of the auxiliary basis; we achieve accuracies on
the order of $10^{-6}~E_\mathrm{h}$ per atom with speedups of one to two orders of magnitude compared
to reference GPW calculations.

While finalizing this work, a preprint by Sun \cite{Sun20arXiv} reported a
similar range-separation idea to accelerate the direct computation of the
four-center Coulomb and the exchange integrals for periodic HF calculations, i.e.~without
density fitting. Therefore, we will also compare our RSGDF to this new method
(referred to as RSJK henceforth) in terms of accuracy and computational cost.

\textit{Theory}. We begin with a brief review of periodic GDF using a basis of
$n_{\tr{AO}}$ symmetry-adapted atomic orbitals (AOs)
    \begin{equation}    \label{eq:TAAO}
        \phi_{\mu}^{\bm{k}}(\bm{r})
            = \sum_{\bm{m}} \me^{\mi \bm{k}\cdot\bm{m}}
            \phi_{\mu}^{\bm{m}}(\bm{r})
    \end{equation}
where $\bm{k}$ is a crystal momentum in the first Brillouin zone, $\bm{m}$ is a
lattice translation vector, and $\phi_{\mu}^{\bm{m}}(\bm{r}) = \phi_{\mu}(\bm{r}-\bm{m})$.
An analogous equation holds for the $n_{\tr{aux}}$ auxiliary atom-centered Gaussian basis
$\chi_P^{\bm{k}}(\bm{r})$. The ERIs are the Coulomb repulsion between pair densities
    \begin{equation}    \label{eq:ERI}
        (\rho_{\mu\nu}^{\bm{k}_1\bm{k}_2} |
        \rho_{\lambda\sigma}^{\bm{k}_3\bm{k}_4})
            = \int_{\Omega}\md\bm{r}_1 \int\md\bm{r}_2\,
            \frac{
                \rho_{\mu\nu}^{\bm{k}_1\bm{k}_2}(\bm{r}_1)
                \rho_{\lambda\sigma}^{\bm{k}_3\bm{k}_4}(\bm{r}_2)
            }
            {r_{12}}
    \end{equation}
where $\Omega$ is the unit cell volume,
$\rho_{\mu\nu}^{\bm{k}_1\bm{k}_2}(\bm{r}) = \phi_{\mu}^{\bm{k}_1*}(\bm{r})
\phi_{\nu}^{\bm{k}_2}(\bm{r})$,
and the four crystal momenta satisfy
$(\bm{k}_1-\bm{k}_2 + \bm{k}_3 - \bm{k}_4) \cdot \bm{m} = 0$ for all $\bm{m}$.
In GDF, the pair densities are approximated by an auxiliary expansion
    \begin{equation}    \label{eq:aux_expansion_rho}
        \rho_{\mu\nu}^{\bm{k}_1\bm{k}_2}(\bm{r})
            \approx
            \sum_{Q}^{n_{\tr{aux}}}
            d_{Q\mu\nu}^{\bm{k}_{1}\bm{k}_2}
            \chi_{Q}^{\bm{k}_{12}}(\bm{r}),
    \end{equation}
with $\bm{k}_{12} = -\bm{k}_1 + \bm{k}_2$. Minimizing the fitting error in some metric $w(r_{12})$
leads to a linear equation for $\mat{d}^{\bm{k}_1\bm{k}_2}$,
    \begin{equation}    \label{eq:DF_eqn}
        \sum_{Q}^{n_{\tr{aux}}} J_{PQ}^{\bm{k}_{12}}
        d_{Q\mu\nu}^{\bm{k}_1\bm{k}_2}
            = V_{P\mu\nu}^{\bm{k}_1\bm{k}_2},
    \end{equation}
where the two- and three-center metric integrals are
    \begin{align}    \label{eq:JPQ}
        J_{PQ}^{\bm{k}}
            &= (\chi_P^{\bm{k}*} | w | \chi_Q^{\bm{k}}), \\
        \label{eq:VPmunu}
        V_{P\mu\nu}^{\bm{k}_1\bm{k}_2}
            &= (\chi_P^{\bm{k}_{12}*} | w | \rho_{\mu\nu}^{\bm{k}_1\bm{k}_2}).
    \end{align}
Once $\{\mat{d}^{\bm{k}_1\bm{k}_2}\}$ are determined, the ERIs can be easily recovered,
    \begin{equation}    \label{eq:ERI_aux}
        (\rho_{\mu\nu}^{\bm{k}_1\bm{k}_2} |
        \rho_{\lambda\sigma}^{\bm{k}_3\bm{k}_4})
            \approx \sum_{P,Q}^{n_{\tr{aux}}} d_{P\mu\nu}^{\bm{k}_1\bm{k}_2}
            (\chi_P^{\bm{k}_{12}} | \chi_Q^{\bm{k}_{34}})
            d_{Q\lambda\sigma}^{\bm{k}_3\bm{k}_4}.
    \end{equation}
The fixed size of the auxiliary Gaussian basis is responsible for a DF error
compared to a calculation without DF (throughout, we will call this the
\textit{accuracy}, to be contrasted with the \textit{precision} with which the
two- and three-center integrals are evaluated for a fixed auxiliary basis).
Although in principle the same is true of GPW, the auxiliary PW basis is
typically grown to achieve arbitrarily accurate results that are free of DF
error.

The computational bottleneck of periodic GDF is due to the three-center
integrals in \cref{eq:VPmunu}, which, when using the long-ranged Coulomb metric $w(r_{12}) =
r_{12}^{-1}$, are expensive to evaluate in real space or reciprocal space.
The current implementation of periodic GDF in PySCF~\cite{Sun17JCP} aims to address this
challenge by introducing a Gaussian charge basis $\{\xi_P^{\bm{k}}\}$ to remove the
charge and multipoles of the auxiliary basis. This splits \cref{eq:VPmunu} into two parts
\begin{equation}    \label{eq:RS_chgbas}
    V_{P\mu\nu}^{\bm{k}_1\bm{k}_2}
        = (\chi_{P}^{\bm{k}_{12}} - \xi_{P}^{\bm{k}_{12}} | w
            | \rho_{\mu\nu}^{\bm{k}_1\bm{k}_2}) +
            (\xi_{P}^{\bm{k}_{12}} | w | \rho_{\mu\nu}^{\bm{k}_1\bm{k}_2}).
\end{equation}
The Gaussian exponents of $\{\xi_P^{\bm{k}}\}$ are optimized so that
the first term in \cref{eq:RS_chgbas} can be evaluated in real space using a lattice summation
and the second term in reciprocal space using \cref{eq:VPmunu_FT}.
Although this yields an improvement over any attempt to evaluate the three-center integrals entirely
in real or reciprocal space, the two separate summations can both be relatively slow to converge.
Our new periodic RSGDF takes a different approach to evaluate the three-center integrals in \cref{eq:VPmunu}. \rv{All techniques introduced below can be readily adapted to the evaluation of the two-center integrals in \cref{eq:JPQ}.}

In RSGDF, we range-separate the Coulomb operator using the error function \cite{Gill96CPL}
$r_{12}^{-1} = w^{\tr{SR}}(r_{12}; \omega) + w^{\tr{LR}}(r_{12}; \omega)$,
    \begin{subequations}    \label{eq:Coulomb_RS}
    \begin{align}
        w^{\tr{SR}}(r_{12}; \omega) &= \frac{\tr{erfc}(\omega r_{12})}{r_{12}} \\
        w^{\tr{LR}}(r_{12}; \omega) &= \frac{\tr{erf}(\omega r_{12})}{r_{12}}
    \end{align}
    \end{subequations}
so that \cref{eq:VPmunu} is split into a short-range (SR) part and a long-range
(LR) part with $\omega$ controlling their relative weights.
We evaluate the LR integrals in reciprocal space,
    \begin{equation}    \label{eq:VPmunu_LR}
        (V_{P\mu\nu}^{\bm{k}_1\bm{k}_2})^{\tr{LR}}_{\omega}
            = 4\pi \sum_{\bm{G}}^{N_{\tr{PW}}}{}^{'}
            \frac{
                \me^{-|\bm{G}+\bm{k}_{12}|^2/4\omega^2}}
            {|\bm{G}+\bm{k}_{12}|^2}
            \tilde{\chi}_{P}^{\bm{k}_{12}}(-\bm{G})
            \tilde{\rho}_{\mu\nu}^{\bm{k}_1\bm{k}_2}(\bm{G}),
    \end{equation}
where $\tilde{\chi}$ and $\tilde{\rho}$ are the Fourier transform of $\chi$
and $\rho$ and the primed summation indicates $\bm{G} \neq
\bm{0}$ for $\bm{k}_1 = \bm{k}_2$~\cite{McClain17JCTC};
\rv{the $\bm{G}=\bm{0}$ term contributes to finite-size errors and is handled on a case-by-case
basis in the subsequent electronic structure calculations and not in the ERIs.}
A relatively small number of PWs are necessary for convergence due to the presence of the
Gaussian damping factor.
The analytical Fourier transform (AFT) is needed for compact auxiliary
orbitals and pair densities, while the fast Fourier transform (FFT) can be used for diffuse ones
\cite{FustiMolnar02JCP,FustiMolnar02JCP2}; we will return to this point later.
The cost of this step is therefore dominated by the AFT of the orbital pair densities
    \begin{equation}    \label{eq:AFT_pairden}
        \tilde{\rho}_{\mu\nu}^{\bm{k}_1\bm{k}_2}(\bm{G})
            = \sum_{\bm{m}}^{N_{\tr{cell}}^{\tr{AFT}}}
            \me^{-\mi\bm{k}_2 \cdot \bm{m}}
            \int \md \bm{r}\,
            \phi_{\mu}^{\bm{0}}(\bm{r}) \phi_{\nu}^{\bm{m}}(\bm{r})
            \me^{-\mi(\bm{k}_{12}+\bm{G})\cdot\bm{r}}.
    \end{equation}
This AFT  has two separate steps: the evaluation of the real-space integrals and
the subsequent contraction of these integrals with phase factors, which
scale as $O(N_k N_{\tr{cell}}^{\tr{AFT}} N_{\tr{PW}} n_{\tr{AO}}^2)$
and $O(N_k^2 N_{\tr{cell}}^{\tr{AFT}} N_{\tr{PW}} n_{\tr{AO}}^2)$, respectively.
Note that the number of unique crystal momentum pair differences grows linearly with $N_k$
and $N_{\tr{cell}}^{\tr{AFT}}$ can be estimated from the orbital overlap.

The SR part can be easily evaluated in real space by lattice summation
    \begin{equation}    \label{eq:VPmunu_SR}
    \begin{split}
        (V_{P\mu\nu}^{\bm{k}_1\bm{k}_2})^{\tr{SR}}_{\omega}
            &= \sum_{\bm{m}\bm{n}}^{N_{\tr{cell}}}
            \me^{-\mi\bm{k}_1\cdot\bm{m}}
            \me^{\mi\bm{k}_2\cdot\bm{n}}
            (V_{P\mu\nu}^{\bm{0},\bm{m}\bm{n}})^{\tr{SR}}_{\omega} \\
            &\hspace{1em} - \frac{\pi}{\Omega \omega^2} S_{P}^{\bm{k}_{12}}
            S_{\mu\nu}^{\bm{k}_1\bm{k}_2} \delta_{\bm{k}_1,\bm{k}_2}
    \end{split}
    \end{equation}
where
\begin{subequations}
\begin{align}
\label{eq:VPmunu_SR_int}
\begin{split}
(V_{P\mu\nu}^{\bm{0},\bm{m}\bm{n}})^{\tr{SR}}_{\omega}
    &= \int_\Omega \md\bm{r}_1 \int\md\bm{r}_2\ \chi_P^{\bm{0}}(\bm{r_1}) w^{\tr{SR}}(r_{12};\omega) \\
        &\hspace{8em} \times \phi_{\mu}^{\bm{m}}(\bm{r_2})\phi_{\nu}^{\bm{n}}(\bm{r_2}),
\end{split}\\
S_{P}^{\bm{k}_{12}} &= \int_{\Omega} \md\bm{r}\, \chi_{P}^{\bm{k}_{12}}(\bm{r}), \\
S_{\mu\nu}^{\bm{k}_1\bm{k}_2} &= \int_{\Omega} \md\bm{r}\, \rho_{\mu\nu}^{\bm{k}_1\bm{k}_2}(\bm{r}),
\end{align}
\end{subequations}
and the summation range $N_{\tr{cell}}$ scales as $O(\omega^{-3})$ for three-dimensional
solids because $\omega^{-1}$ is the decay length of the SR potential.
The second term in \cref{eq:VPmunu_SR} cancels the $\bm{G}=\bm{0}$ component
of the first term. With proper integral screening, only $O(N_{\tr{cell}})$ terms
contribute significantly to the double lattice summation to achieve a finite
precision $\epsilon$ (see Supplementary Material for a detailed derivation).
Therefore, the costs scale as $O(N_{\tr{cell}}
n_{\tr{aux}} n_{\tr{AO}}^2)$ for the evaluation of real-space integrals in \cref{eq:VPmunu_SR_int}
and $O((N_k^2 N_{\tr{cell}} + N_k N_{\tr{cell}}^2) n_{\tr{aux}} n_{\tr{AO}}^2)$ for the
double phase factor contraction in \cref{eq:VPmunu_SR}.

    \begin{figure}[b]
        \centering
        \includegraphics[scale=0.4]{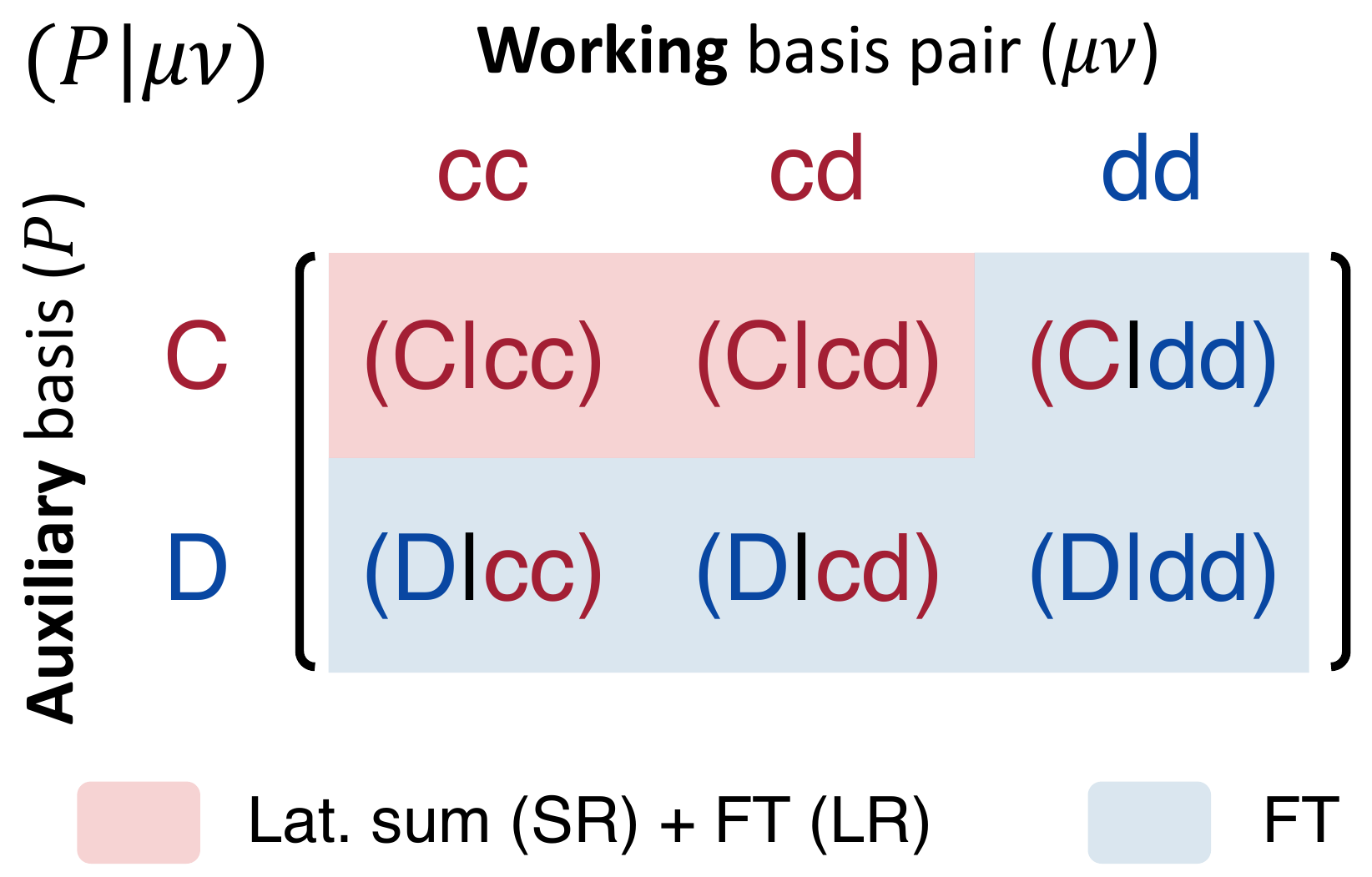}
        \caption{Schematic illustration of how different components of the
three-center integrals [\cref{eq:VPmunu}] are evaluated in RSGDF. Both the
auxiliary and the AO bases are split into a compact (``C/c'') set and diffuse
(``D/d'') set. The compact integrals (shaded in red) are range separated to yield
short-range (SR) and long-range (LR) contributions that are evaluated in real space
(by lattice summation)
and in reciprocal space (by Fourier transforms), respectively;
the diffuse integrals (shaded in blue) are evaluated entirely in reciprocal
space using Fourier transforms. Note that the column corresponding to
``dc''-type pair densities is similar to the ``cd''-column and hence omitted for
simplicity.}
        \label{fig:illustration}
    \end{figure}

    \begin{figure*}[t]
        \centering
        \includegraphics[scale=0.5]{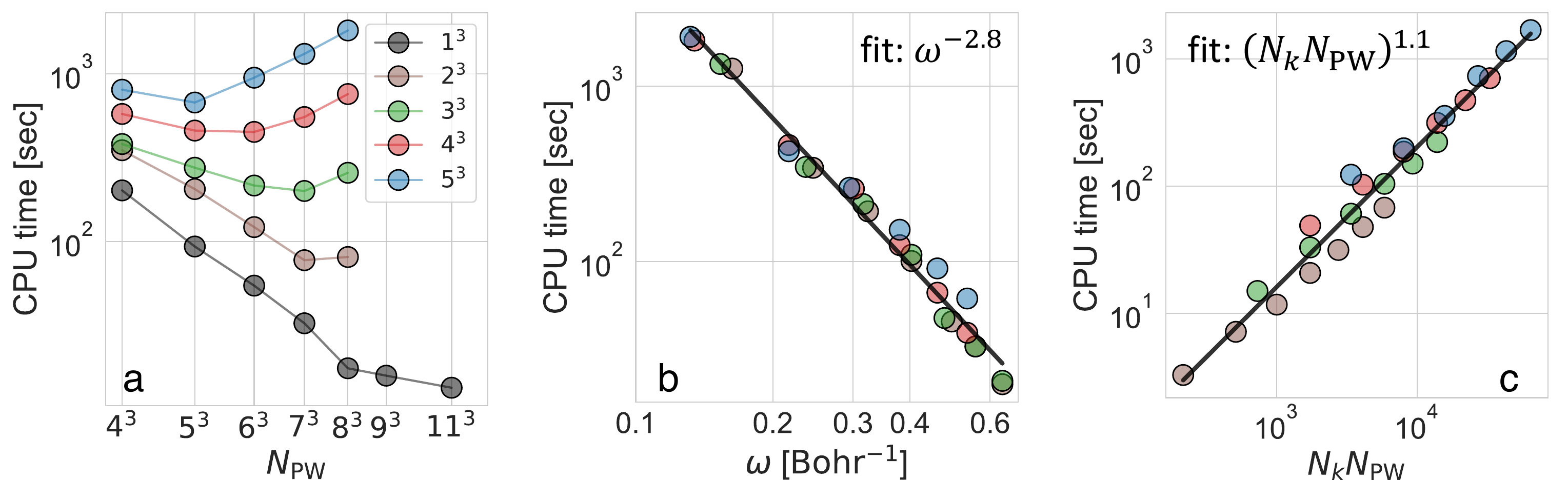}
        \caption{Timing of RSGDF for diamond/cc-pVDZ using $N_k=1^3$ to $5^3$.
(a) Total DF initialization CPU time as a function of the number of PWs
used to compute the LR integrals [\cref{eq:VPmunu_FT,eq:VPmunu_LR}]. (b) CPU
time for computing the SR integrals [\cref{eq:VPmunu_SR}] as a function of
$\omega$ for all $k$-point meshes except for $N_k = 1^3$.  (c) CPU time for computing
the LR integrals [\cref{eq:VPmunu_FT,eq:VPmunu_LR}] as a function of $N_k
N_{\tr{PW}}$ for all $k$-point meshes except for $N_k = 1^3$. For (b) and (c), the black
lines are power law fits to all data points leading to exponents as shown.}
        \label{fig:scaling}
    \end{figure*}

Since $N_{\tr{cell}}^{\tr{AFT}}$ and $N_{\tr{cell}}$ can be as large as $10^4$,
the phase factor contractions in both \cref{eq:AFT_pairden,eq:VPmunu_SR} would account
for most of the computational cost (except for very small $k$-point meshes).
However, if the $k$-points are sampled from a uniform (e.g.~Monkhorst-Pack \cite{Monkhorst76PRB}) mesh
that includes the $\Gamma$ point, then
the phase factors satisfy
$\me^{\mi \bm{k}\cdot\bm{m}} = \me^{\mi \bm{k}\cdot(\bm{\tilde{m}} + \bm{M})}
= \me^{\mi \bm{k}\cdot\bm{\tilde{m}}}$,
where $\bm{\tilde{m}}$ is inside the Born-von Karman supercell and $\bm{M}$ is a lattice translation vector of
the Born-von Karman supercell.
For example, the summation in \cref{eq:VPmunu_SR} can then be rewritten as
    \begin{equation}    \label{eq:BvK_latsum_LR}
        \sum_{\bm{\tilde{m}} \bm{\tilde{n}}}^{N_k}
        \me^{-\mi \bm{k}_1\cdot\bm{\tilde{m}}}
        \me^{\mi \bm{k}_2\cdot\bm{\tilde{n}}}
        \sum_{\bm{m}\to\bm{\tilde{m}},\bm{n}\to\bm{\tilde{n}}}
        (V_{P\mu\nu}^{\bm{0},\bm{m}\bm{n}})^{\tr{SR}}_{\omega},
    \end{equation}
where the phase factor contraction now costs $O(N_k^3 n_{\tr{aux}} n_{\tr{AO}}^2)$;
a similar treatment for \cref{eq:AFT_pairden} gives $O(N_k^3 N_{\tr{PW}}
n_{\tr{AO}}^2)$ cost for the phase factor contraction.
This process
significantly reduces
the total cost of these contractions so that they are subdominant
(at least for moderately sized $k$-point meshes where $N_k$ is much smaller than $N_{\tr{cell}}$ or
$N_{\tr{cell}}^{\tr{AFT}}$).
The remaining cost-determining steps are the real-space integral
evaluations in \cref{eq:AFT_pairden,eq:VPmunu_SR_int}, which as a reminder scale
as $O(N_k N_{\tr{cell}}^{\tr{AFT}}N_{\tr{PW}} n_{\tr{AO}}^2)$ and $O(N_{\tr{cell}} n_{\tr{aux}} n_{\tr{AO}}^2)$,
respectively. Note that the cost of these expensive steps is no worse than linear in the
number of auxiliary Gaussian basis functions.

The algorithm described so far yields significant performance improvements over
existing periodic GDF schemes.  We have identified an additional minor improvement,
motivated by the observation that
the real-space lattice summations needed for the SR part
are slow to converge because of diffuse orbitals, i.e.\ those with small Gaussian
exponents (recall that the FFT can be used in the LR part for diffuse orbitals).
Therefore, we split both the auxiliary
and the AO bases into a compact (``C/c'', upper case for auxiliary) and a diffuse
(``D/d'') set based on a cutoff $\alpha^{\tr{cut}}$ for the primitive Gaussian
exponents.
Similar ideas of compact and diffuse
basis splitting have also been explored by Pulay and co-workers for molecular
calculations. \cite{FustiMolnar02JCP2}
This leads us to six types of three-center integrals as shown in
\cref{fig:illustration}.
Four of the integral types have either a diffuse bra or a diffuse ket (shaded in blue;
note that both ``cc'' and ``cd'' are compact) and can thus be readily evaluated in reciprocal space
using a relatively small PW basis \cite{FustiMolnar02JCP2} \textit{without range separation},
    \begin{equation}    \label{eq:VPmunu_FT}
        V_{P\mu\nu}^{\bm{k}_1\bm{k}_2}
            = 4\pi \sum_{\bm{G}}^{N_{\tr{PW}}}{}^{'} \frac{
                \tilde{\chi}_{P}^{\bm{k}_{12}}(-\bm{G})
                \tilde{\rho}_{\mu\nu}^{\bm{k}_1\bm{k}_2}(\bm{G})
            }
            {|\bm{G}+\bm{k}_{12}|^2}.
    \end{equation}
To summarize, in RSGDF, we evaluate these four integral types directly in reciprocal space
and the remaining two integral types with compact bra and ket using the range separation
scheme defined above.
Note that the expensive AFTs of the compact auxiliary orbitals and pair densities can be calculated
once and used in the evaluation of both integral types.
In practice, a large majority of orbitals are defined as compact and so we find that this
separation of orbitals speeds up
our calculations by a factor of two or less compared to a direct application of RSGDF
for all orbitals.

While the above presentation suggests a computational scaling that is linear in $N_k$ for typical
mesh densities, the actual scaling is complicated by the choice of the
parameters, $N_{\tr{PW}}$, $\omega$, and $\alpha^{\tr{cut}}$, which we discuss more below.
Empirically we find that the optimal choice of $N_{\tr{PW}}$ scales as $O(N_k^{-1/2})$ (Fig.\ S3),
and the overall cost of RSGDF
scales roughly as $O(N_k^{0.8})$ for all the systems tested in this work (Fig.\ S6).
However, this sublinear scaling only holds for small $N_k$, because
$N_{\tr{PW}}$ eventually reaches a minimum value.
Beyond that point, the AFTs needed for the LR part dominate
the cost and we expect a linear scaling of RSGDF with $N_k$, at least for medium-sized $k$-point meshes.

\textit{Computational details}. We implemented RSGDF as presented above in a
local version of the PySCF software package \cite{Sun18WIRCMS}.  We test its
performance in terms of precision, accuracy, and computational efficiency using
three simple three-dimensional solids: diamond, MgO, and LiF. For diamond, we perform
all-electron calculations using the cc-pVDZ basis \cite{Dunning89JCP}; for MgO and LiF, we use
GTH pseudopotentials \cite{Goedecker96PRB,Hartwigsen98PRB} and the corresponding GTH-DZVP basis \cite{VandeVondele05CPC}.
We use the cc-pVDZ-jkfit basis \cite{Weigend02PCCP} and the even-tempered basis
(ETB) generated with a progression factor $\beta = 2.0$ for the auxiliary
expansion of the cc-pVDZ and the GTH-DZVP bases, respectively.

We compare RSGDF to GDF \cite{Sun17JCP}, GPW (called FFTDF in PySCF)
\cite{VandeVondele05CPC}, and RSJK \cite{Sun20arXiv}, as implemented in PySCF.
For RSJK, we use eq.\ (23) in ref.\ \onlinecite{Sun20arXiv} to determine an
appropriate $\omega$ for a given system and $k$-point mesh.  For RSGDF, we
manually test a range of $N_{\tr{PW}}$ and for each we determine the maximum
$\omega$ and $\alpha^{\tr{cut}}$ that guarantee precision $\epsilon$ in all
integrals; future work will focus on the automated selection of these
parameters.  As a general trend, using a larger PW basis slows down the LR part
by increasing the number of expensive real-space integrations to be performed in
\cref{eq:AFT_pairden} but accelerates the SR part by allowing larger values for
$\omega$ and $\alpha^{\tr{cut}}$; a smaller PW basis has the opposite effect.
Unless otherwise mentioned, all calculations are run with a target precision of
$\epsilon = 10^{-8}$ a.u.\ for integral evaluation, which is the default setting
for production-level periodic calculations in PySCF.
\rv{The finite-size error of the HF exchange energy is corrected with a Madelung
constant, which yields $O(N_k^{-1})$ convergence to the thermodynamic limit~\cite{Paier06JCP,Broqvist09PRB,Sundararaman13PRB} \rvv{(see Sec.\ S3 for details)}; other possibilities exist~\cite{Gygi86PRB,Spencer08PRB,Guidon09JCTC,Sundararaman13PRB} but would require modification of the DF algorithm.}
All timing data reported
below are the CPU time recorded using a single CPU core (Intel Xeon Gold 6126
2.6 GHz) with $16$ GB of memory and $100$ GB of disk space except for GPW which
requires larger memory for $N_k \geq 4^3$.  The current implementations of GDF
and RSGDF are not integral-direct, meaning that we solve \cref{eq:DF_eqn} only
once and save the coefficients $\{\mat{d}^{\bm{k}_1\bm{k}_2}\}$ to disk for
later use; this step is called ``DF initialization'' below and requires $O(N_k^2
n_{\tr{aux}} n_{\tr{AO}}^2)$ disk space which limits our calculations to a
maximum $k$-point mesh of $N_k=5^3$ for all three systems.  The other two
methods, GPW and RSJK, are both implemented in an integral-direct manner and
hence require little disk space.  An integral-direct implementation of RSGDF
tailored for specific applications will be presented in future work.

\textit{Results and discussion}. We first verify our scaling analysis of the CPU cost of
RSGDF.  In \cref{fig:scaling}a, we show the RSGDF initialization time as a
function of $N_{\tr{PW}}$ for diamond using $N_k=1^3$ to $5^3$ $k$-points (recall that
all calculations achieve the same target precision).  The
optimal $N_{\tr{PW}}$ -- identified as the minimum on each curve -- indeed
decreases with $N_k$ as $O(N_k^{-1/2})$ (the fitted exponent is about $-0.44$;
see Fig.\ S3).  The inverse cubic dependence of the SR time on $\omega$ is
verified in \cref{fig:scaling}b, and the linear scaling of the LR time with
$N_{k} N_{\tr{PW}}$ is verified in \cref{fig:scaling}c.  Similar
results are observed for the other two systems (Figs.\ S1 and S2), although the
optimal values of $N_{\tr{PW}}$ for a given $N_k$ vary slightly from system to
system.  We leave the automatic determination of the optimal $N_{\tr{PW}}$ to a
future work as it requires a more careful calibration.
In what follows, we will simply use the manually optimized
values from \cref{fig:scaling}a and Figs.\ S1a and S2a (summarized in Tab.\ S1).

The different choices of $N_{\tr{PW}}$ (and hence $\omega$ and $\alpha^{\tr{cut}}$) in RSGDF do not
cause any inconsistency in the computed energies. As shown in Fig.\ S4, the
converged RSGDF HF energies differ from the GDF results by less than $10^{-7}~E_{\tr{h}}$
for diamond and MgO and about $10^{-6}~E_{\tr{h}}$ for LiF for all data points shown
in \cref{fig:scaling} and Figs.\ S1 and S2. We attribute the larger deviation
observed for LiF to the linear dependency found in the auxiliary basis.
Nonetheless, these deviations are acceptable as they are at least one order of
magnitude smaller than the error introduced by DF itself (\textit{vide infra}).
Beyond the HF energy, we have also verified that the electron correlation energy of diamond computed
with RSGDF using the second order M{\o}ller-Plesset perturbation theory \cite{Moller34PR}
agrees with the GDF results to better than $10^{-8}~E_{\tr{h}}$ for all $k$-point meshes tested
(Tab.~S2). These observations confirm that the algorithmic developments in
RSGDF cause negligible precision loss compared to the original implementation of
GDF.

    \begin{figure*}[t]
        \centering
        \includegraphics[scale=0.4]{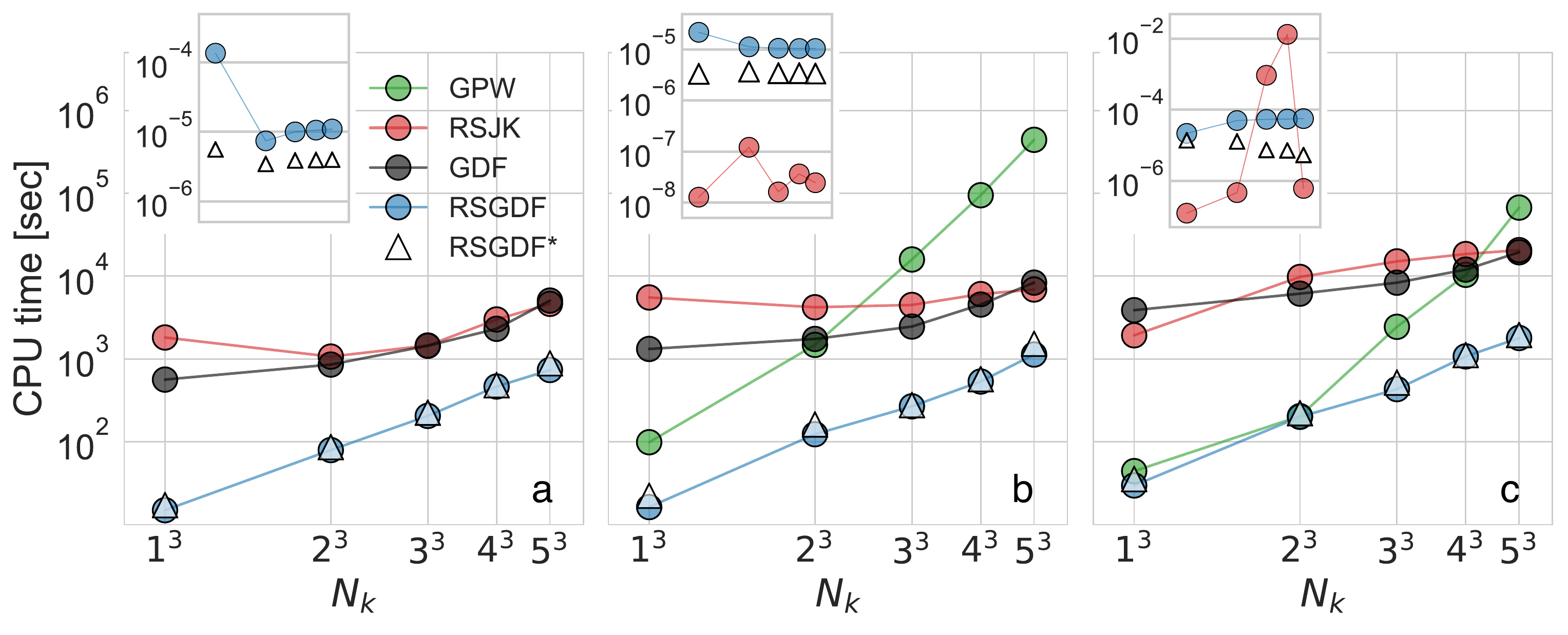}
        \caption{CPU time (per SCF cycle) for computing the Coulomb and the
exchange integrals in a HF calculation for (a) diamond/cc-pVDZ, (b)
MgO/GTH-DZVP, and (c) LiF/GTH-DZVP using GPW (green), RSJK (red), GDF (grey),
and RSGDF (blue) to handle the ERIs. For GDF and RSGDF, the DF initialization
time is included. For RSGDF, results using a larger auxiliary basis are also included
(RSGDF*, white triangles). \rv{Precise timing data are given in Table S3.}
        Insets show the deviation of the RSGDF HF energies from RSJK for
diamond (a) and deviations of the RSGDF and the RSJK HF energies from GPW for MgO (b)
and LiF (c), where the x-axis is $N_k$ and the y-axis is in $E_{\tr{h}}$.}
        \label{fig:df_vs_nodf}
    \end{figure*}

Next, we study the computational efficiency of RSGDF. In \cref{fig:df_vs_nodf},
we plot the per-SCF-cycle time as a function of $N_k$ for computing the Coulomb
and the exchange integrals in a HF calculation for all three systems using four
different methods to handle the ERIs: GPW (green), RSJK (red), GDF (grey), and
RSGDF (blue). Since the first two are implemented in an integral-direct fashion,
we include the DF initialization time for GDF and RSGDF to enable a fair
comparison.

We first compare GDF and RSGDF, which both compute the three-center integrals
through a SR part in real space and a LR part in reciprocal space.  Although
both methods exhibit a similar sublinear scaling with $N_k$ at large $N_k$, only
the GDF timings plateau at small $N_k$.  This difference arises from the
adjustment of the PW basis size for each $N_k$ and for each system (Tab.~S1),
which ultimately balances the SR-LR cost in RSGDF as analyzed above (Fig.\ S5).
More importantly, the algorithmic optimizations developed for RSGDF in this work
significantly reduce its computational cost and lead to speedups of one to
two orders of magnitude over the previous GDF for all three systems studied here.

We next compare RSGDF with the two other methods without DF error, i.e.~GPW and RSJK.
The GPW timing shows the characteristic $O(N_k^2)$ scaling of computing exact
exchange starting from $N_k = 2^3$ and is 40 to 400 times slower
than RSGDF for the largest $N_k$ tested here.
The very high cost of GPW for MgO is caused by the compact primitive
Gaussians in the $2s$ and $2p$ orbitals of Mg, which require $83^3$ PWs to reach
the target precision of $10^{-8}$ (cf.\ $51^3$ for LiF).
We emphasize that lowering the precision requirement for GPW (hence lowering
$N_{\tr{PW}}$) only moderately reduces the cost due to the $O(N_{\tr{PW}}\ln
N_{\tr{PW}})$ scaling of FFT.  For example, using $\epsilon = 10^{-6}$ and $10^{-4}$ requires
$75^3$ and $66^3$ PWs for MgO and reduces the cost by only factors of $1.4$ and
$2.1$, respectively.

The RSJK timings are similar to those of GPW; although they show a slightly
weaker dependence on $N_k$, the precise scaling is unclear.  This peculiar
$N_k$-dependence of RSJK arises from a significant SR-LR cost unbalance
(Fig.\ S5) and suggests a breakdown of eq.\ (23) in ref.\ \onlinecite{Sun20arXiv} for
determining the optimal $\omega$ for RSJK.  Despite this, RSJK still achieves
computational efficiency similar to GDF and much higher than GPW for moderately
sized $k$-point meshes, which is remarkable given that RSJK does not use DF.

Finally, we examine the accuracy of the HF energies computed by RSGDF, which is
a combination of the DF error due to the auxiliary basis set and the precision
error when calculating the matrices in \cref{eq:JPQ,eq:VPmunu}.
The RSJK results are used as the benchmark for diamond and the GPW results are used
as the benchmark for MgO and LiF.
To probe the possibility of achieving higher accuracy with
DF, we also include results for RSGDF using a larger auxiliary basis (denoted by
``RSGDF*'' in \cref{fig:df_vs_nodf}); we use the cc-pVTZ-jkfit basis for diamond
and an ETB whose size is about 1.25 times larger (obtained by using a smaller
$\beta$) for MgO and LiF.  The per-atom errors of the converged HF energies are
plotted in the insets of \cref{fig:df_vs_nodf}.

With the default auxiliary basis, the error introduced by RSGDF is about
$10^{-5}~E_{\tr{h}}$ for diamond and MgO and about $10^{-4}~E_{\tr{h}}$ for LiF;
these errors are typical for DF-based HF calculations
\cite{Burow09JCP,Patterson20JCP}.  Using the slightly larger auxiliary basis
(RSGDF*) reduces the error by a factor of three or more and, most remarkably,
requires little extra work as can be seen from the nearly identical timings of
RSGDF and RSGDF* in \cref{fig:df_vs_nodf}.  \rv{This is because the LR part is dominated by the AFTs of the orbital pair densities, \cref{eq:AFT_pairden}, whose cost is independent of $n_{\tr{aux}}$, while the SR part scales linearly with $n_{\tr{aux}}$.}  By contrast, the accuracy of RSJK is in general very high
($10^{-7}~E_{\tr{h}}$ or less), but relatively large errors of about
$10^{-2}~E_{\tr{h}}$ are also observed for certain $k$-point meshes (inset of
\cref{fig:df_vs_nodf}c); the accuracy loss in the latter cases is likely due to
an inaccurate integral screening, as tightening the precision to $10^{-10}$
reduces the error to about $10^{-5}~E_h$ for the calculation of LiF using $4^3$
$k$-points.  These results demonstrate that RSGDF provides an extremely
cost-effective approach to calculating accurate HF energies in periodic systems.

\textit{Conclusion}. To summarize, we have presented an efficient scheme that
uses range separation for Gaussian density fitting (RSGDF) for periodic systems.
The computational scaling is analyzed to be sublinear with $N_k$ for small
$k$-point meshes and linear for medium-sized ones.  With all-electron and
pseudopotential-based numerical results on a few three-dimensional solids, we
verified the scaling of RSGDF and showed that it achieves about $10$-fold
speedups over the previously developed GDF with little precision loss.  The
error introduced by RSGDF is about $10^{-5}~E_{\tr{h}}$ with default auxiliary
basis sets and can be systematically reduced by increasing the size of the
auxiliary basis with little extra work.

The primary purpose of the current integral-indirect implementation of RSGDF is
to speed up Hartree-Fock (and hybrid density functional theory) calculations for
a given Gaussian basis and auxiliary basis.  Motivated by the excellent
performance seen in these preliminary calculations, we are currently working on
the automatic determination of the optimal $N_{\tr{PW}}$, $\omega$, and
$\alpha^{\tr{cut}}$.  Looking forward, the fast integral construction enabled by
RSGDF encourages the development of integral-direct algorithms tailored to
specific tasks such as the evaluation of exact exchange \cite{Wang20JCP}, the
ERI orbital transformation \cite{Maschio07PRB,Usvyat07PRB}, and post-HF
calculations \cite{Usvyat07PRB,Luenser17JCTC}.  Such integral-direct methods
would significantly reduce the high memory footprint currently required by
post-HF calculations on periodic systems with Gaussian basis sets.

\section*{Supplementary material}

See the supplementary material for (i) RSGDF initialization time for different
choices of $N_{\tr{PW}}$ and $\omega$ for MgO and LiF; (ii) optimal choices of
$N_{\tr{PW}}$ for $N_k$ from $1^3$ to $5^3$ and different systems; (iii)
difference of the HF energies computed using RSGDF and GDF; (iv) SR and LR
component time of RSGDF, GDF, and RSJK for varying size of $k$-point meshes;
(v) $N_k$-scaling of RSGDF and GDF for the
timing data shown in \cref{fig:df_vs_nodf}; (vi) values of the parameters
needed by RSGDF, GDF, RSJK, and GPW; (vii) comparison of the MP2 correlation
energies computed using RSGDF and GDF for diamond; \rv{(viii) CPU time for computing the Coulomb and exchange integrals per SCF cycle by RSGDF, GDF, RSJK, and GPW;} \rvv{(ix) Details of the treatment of exchange divergence;} (x) derivations of the conditions for prescreening the double lattice summation in \cref{eq:VPmunu_SR}.

\section*{Acknowledgements}

HY thanks Dr.\ Qiming Sun and Dr.\ Xiao Wang for helpful discussions.
This work was supported by the National Science Foundation under Grant No.\
OAC-1931321.  We acknowledge computing resources from Columbia University's
Shared Research Computing Facility project, which is supported by NIH Research
Facility Improvement Grant 1G20RR030893-01, and associated funds from the New
York State Empire State Development, Division of Science Technology and
Innovation (NYSTAR) Contract C090171, both awarded April 15, 2010. The Flatiron
Institute is a division of the Simons Foundation.

\section*{Data availability statement}
The data that support the findings of this study are available from the
corresponding author upon reasonable request.

\bibliography{refs}

\raggedbottom

\end{document}


\maketitle

    \tableofcontents

    \vspace{3em}

    \hspace{2em}Note: figures and equations appearing in the main text will be referred as
``Fig.\ Mxxx'' and ``Eq.\ Mxxx'' in this Supplementary Material document.

    \clearpage

    \section{Supplementary figures}

    \begin{figure}[!h]
        \centering
        \includegraphics[width=1.0\linewidth]{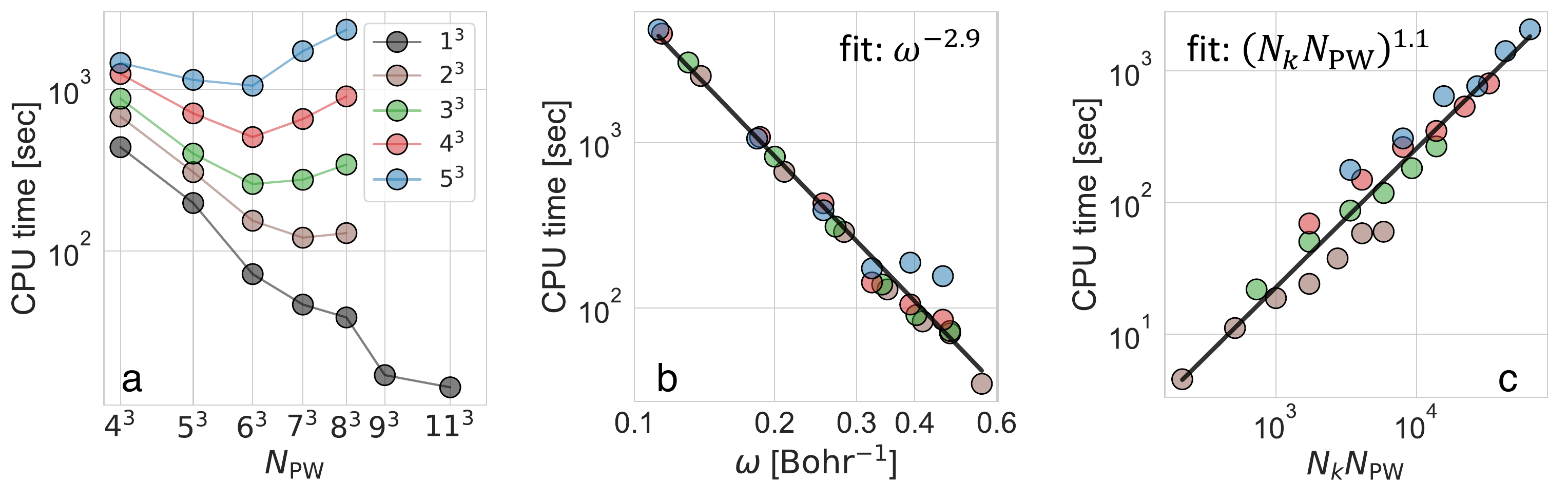}
        \caption{Same plot as Fig.\ M1 for MgO/GTH-DZVP.}
        \label{fig:MgO_scaling}
    \end{figure}

    \begin{figure}[!h]
        \centering
        \includegraphics[width=1.0\linewidth]{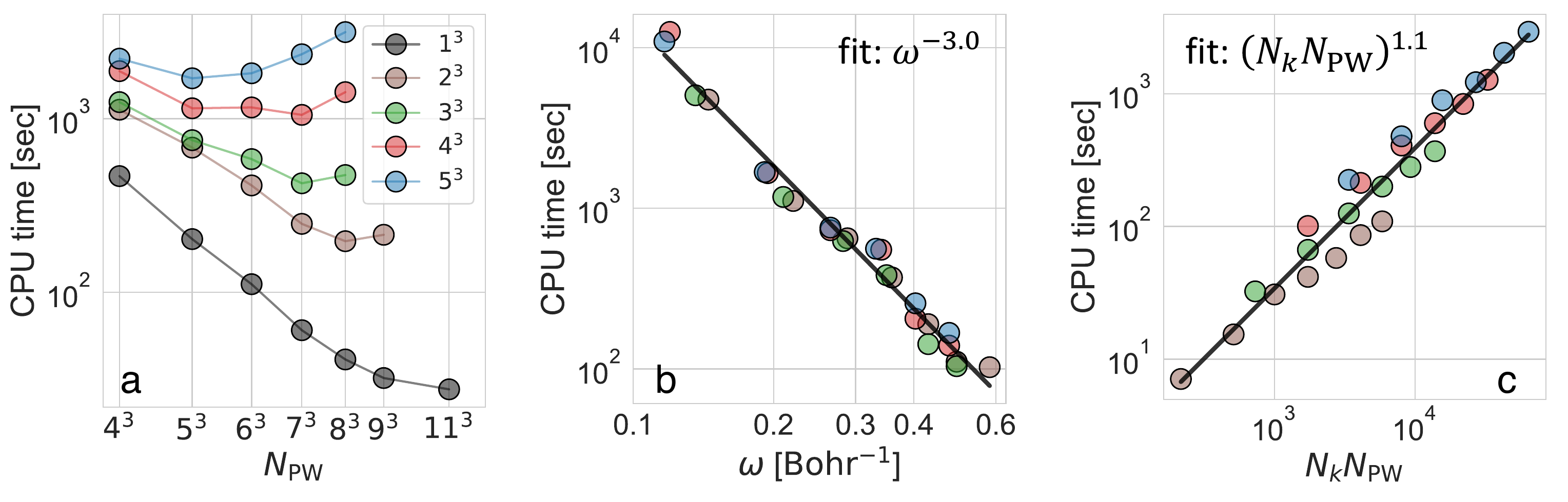}
        \caption{Same plot as Fig.\ M1 for LiF/GTH-DZVP.}
        \label{fig:LiF_scaling}
    \end{figure}

    \begin{figure}[!h]
        \centering
        \includegraphics[width=0.35\linewidth]{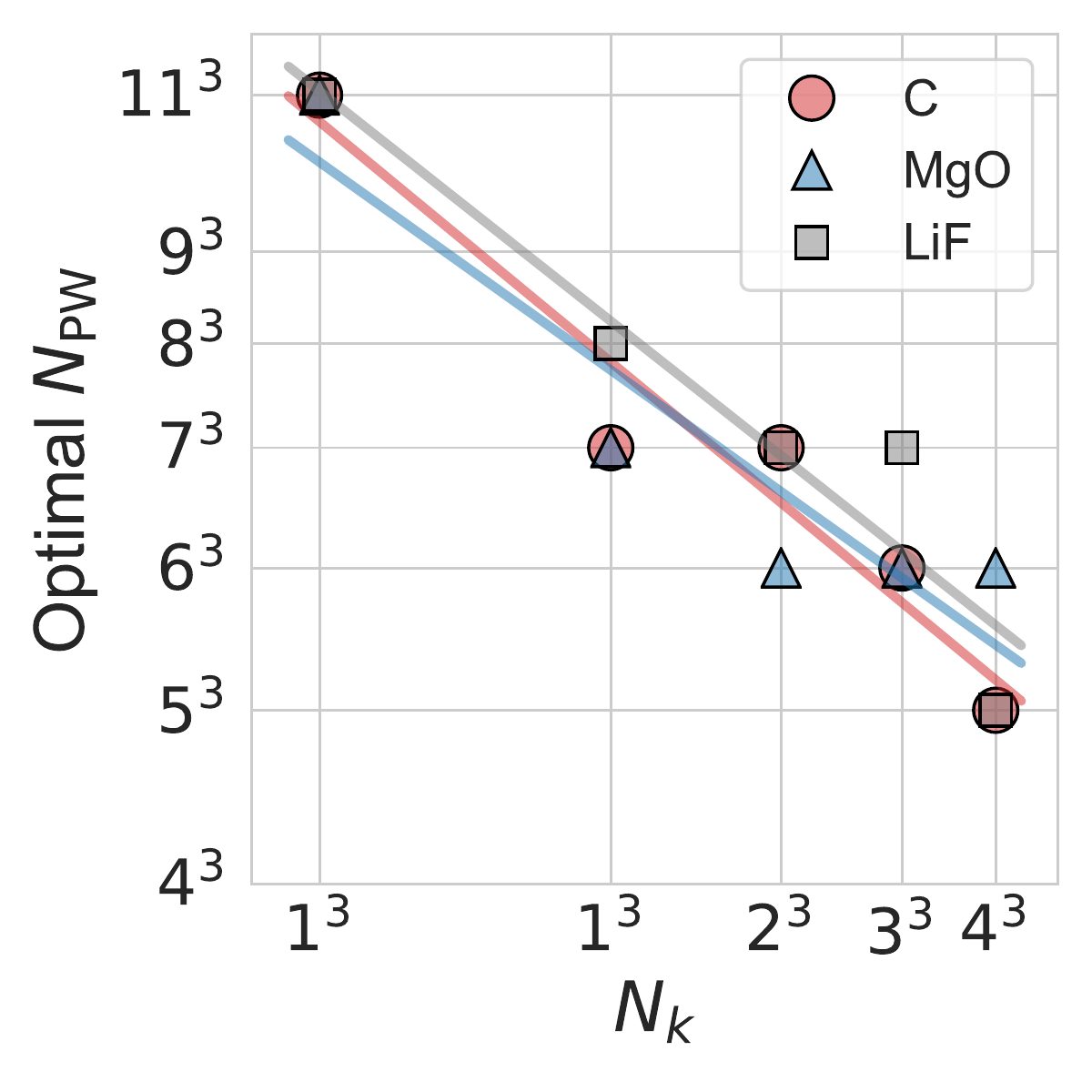}
        \caption{Optimal choice of $N_{\tr{PW}}$ (obtained from Fig.\ M1 and \cref{fig:MgO_scaling,fig:LiF_scaling}) plotted as a function of $N_k$. The estimated exponents are $-0.44$ (C), $-0.38$ (MgO), and $-0.43$ (LiF).}
        \label{fig:npw_nk_fit}
    \end{figure}

    \begin{figure}[!h]
        \centering
        \includegraphics[width=1.0\linewidth]{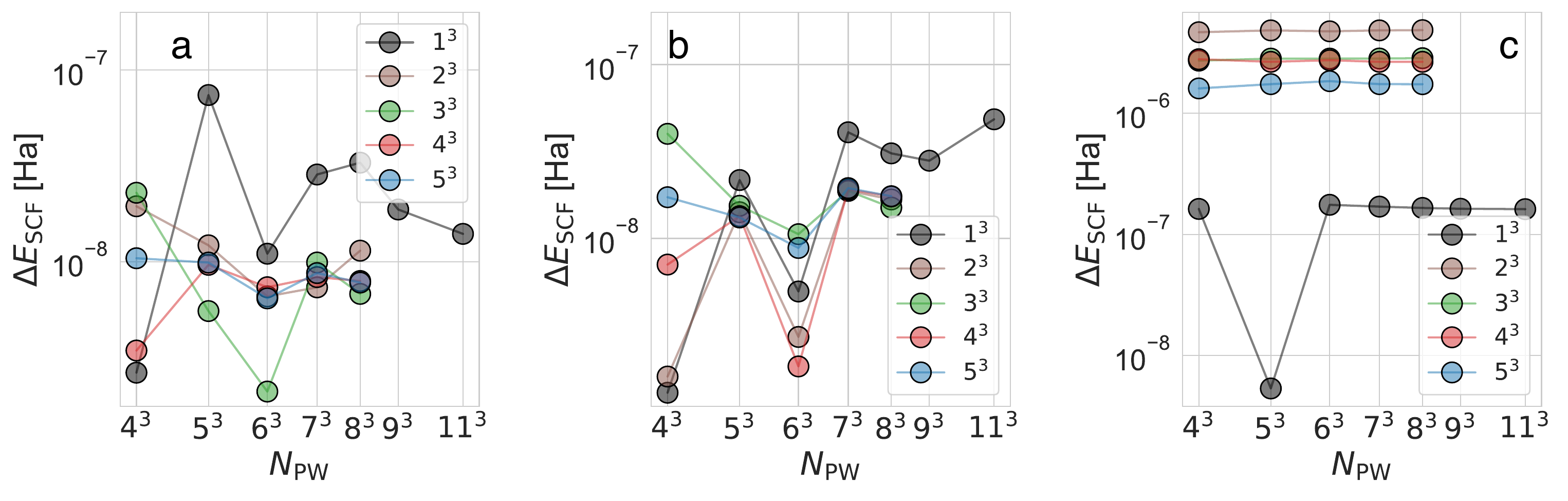}
        \caption{Absolute deviation between the converged HF energies computed using RSGDF and GDF for (a) diamond/cc-pVDZ, (b) MgO/GTH-DZVP, and (c) LiF/GTH-DZVP using $1^3$ to $5^3$ $k$-points.}
        \label{fig:HF_error}
    \end{figure}

    \begin{figure}[!h]
        \centering
        \includegraphics[width=0.9\linewidth]{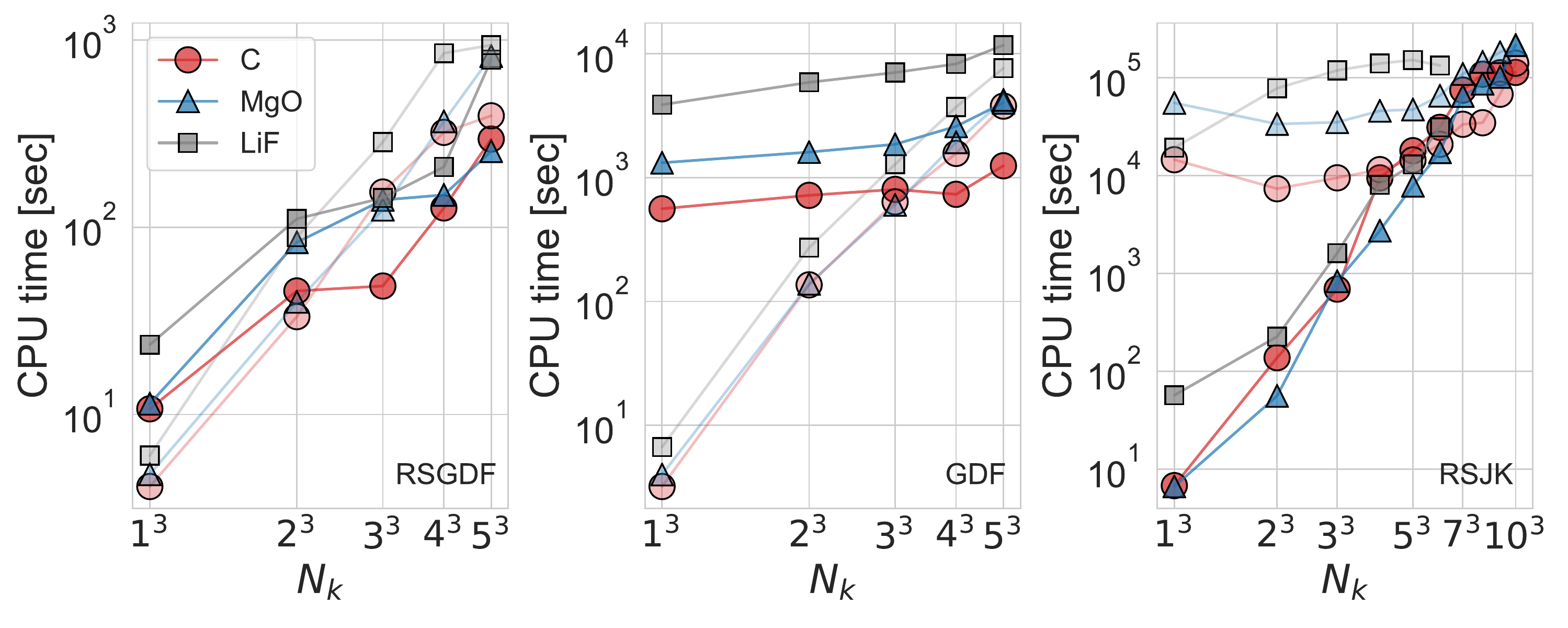}
        \caption{Component timing data (SR: darker; LR: lighter) plotted as a function of $N_k$ for RSGDF (left), GDF (middle), and RSJK (right).}
        \label{fig:SR_vs_LR}
    \end{figure}

    \begin{figure}[!h]
        \centering
        \includegraphics[width=0.9\linewidth]{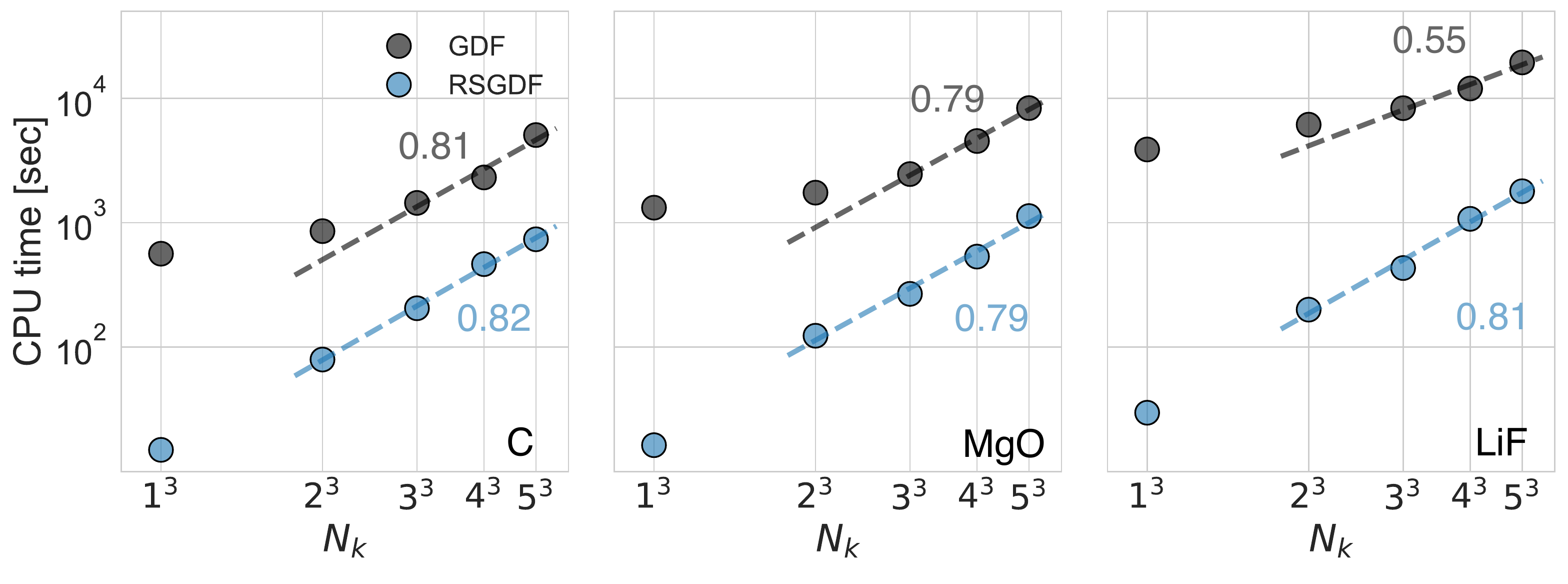}
        \caption{Same data for GDF and RSGDF as plotted in Fig.\ M3 with power-law fitting ($t = a N_k^{\gamma}$) where the exponent $\gamma$ is displayed in the figure. For RSGDF, the last four points are used, while for GDF, the last three points are used.}
        \label{fig:scaling}
	\end{figure}

    \clearpage

    \section{Supplementary tables}

    \begin{table}[!h]
        \centering
        \caption{Parameters for each method used in this work. RSGDF/RSJK: $\omega$ and $N_{\tr{PW}}$. GDF: charge basis exponent $\alpha_{\tr{chg}}$ and $N_{\tr{PW}}$. GPW: $N_{\tr{PW}}$.}
        \label{tab:parameters}
        \begin{tabular}{cccccccc}
            \toprule
            \multicolumn{8}{c}{RSGDF}   \\
            \midrule
            \multirow{2}*{$N_k$} &
                    \multirow{2}*{$N_{\bm{k}_{12}}$} &
                    \multicolumn{2}{c}{Diamond} &
                    \multicolumn{2}{c}{MgO} &
                    \multicolumn{2}{c}{LiF} \\
            \cmidrule(lr){3-4} \cmidrule(lr){5-6} \cmidrule(lr){7-8}
            & & $\omega$ & $N_{\tr{PW}}$ &
                $\omega$ & $N_{\tr{PW}}$ &
                $\omega$ & $N_{\tr{PW}}$    \\
            \midrule
            $1^3$ & $1$ &
                    $0.792570$ & $11^3$ &
                    $0.671035$ & $11^3$ &
                    $0.707182$ & $11^3$ \\
            $2^3$ & $14$ &
                    $0.494208$ & $7^3$ &
                    $0.415922$ & $7^3$ &
                    $0.494208$ & $8^3$ \\
            $3^3$ & $63$ &
                    $0.476279$ & $7^3$ &
                    $0.340426$ & $6^3$ &
                    $0.429530$ & $7^3$ \\
            $4^3$ & $172$ &
                    $0.379822$ & $6^3$ &
                    $0.323641$ & $6^3$ &
                    $0.403150$ & $7^3$ \\
            $5^3$ & $365$ &
                    $0.294591$ & $5^3$ &
                    $0.323641$ & $6^3$ &
                    $0.264873$ & $5^3$ \\
            \bottomrule
        \end{tabular}

        \vspace{1em}

        \begin{tabular}{ccccccc}
            \toprule
            \multicolumn{7}{c}{RSJK}   \\
            \midrule
            \multirow{2}*{$N_k$} &
                    \multicolumn{2}{c}{Diamond} &
                    \multicolumn{2}{c}{MgO} &
                    \multicolumn{2}{c}{LiF} \\
            \cmidrule(lr){2-3} \cmidrule(lr){4-5} \cmidrule(lr){6-7}
            &
                $\omega$ & $N_{\tr{PW}}$ &
                $\omega$ & $N_{\tr{PW}}$ &
                $\omega$ & $N_{\tr{PW}}$    \\
            \midrule
            $1^3$ & $5.664491$ & $69^3$ & $5.768690$ & $83^3$ & $3.700892$ & $51^3$ \\
            $2^3$ & $1.888164$ & $23^3$ & $2.224073$ & $32^3$ & $2.902661$ & $40^3$ \\
            $3^3$ & $1.149317$ & $14^3$ & $1.320543$ & $19^3$ & $1.741596$ & $24^3$ \\
            $4^3$ & $0.820941$ & $10^3$ & $0.973032$ & $14^3$ & $1.233631$ & $17^3$ \\
            $5^3$ & $0.656753$ & $8^3$ & $0.764525$ & $11^3$ & $1.015931$ & $14^3$ \\
            $6^3$ & $0.574659$ & $7^3$ & $0.625521$ & $9^3$ & $0.798232$ & $11^3$ \\
            $7^3$ & $0.492564$ & $6^3$ & $0.556018$ & $8^3$ & & \\
            $8^3$ & $0.410470$ & $5^3$ & $0.486516$ & $7^3$ & & \\
            $9^3$ & $0.410470$ & $5^3$ & $0.417014$ & $6^3$ & & \\
            $10^3$ & $0.410470$ & $5^3$ & $0.347511$ & $5^3$ & & \\
            \bottomrule
        \end{tabular}

        \vspace{1em}

        \begin{tabular}{ccccccc}
            \toprule
            \multicolumn{7}{c}{GDF}  \\
            \midrule
            \multirow{2}*{$N_k$} &
                    \multicolumn{2}{c}{Diamond} &
                    \multicolumn{2}{c}{MgO} &
                    \multicolumn{2}{c}{LiF} \\
            \cmidrule(lr){2-3} \cmidrule(lr){4-5} \cmidrule(lr){6-7}
            &
                $\alpha_{\tr{chg}}$ & $N_{\tr{PW}}$ &
                $\alpha_{\tr{chg}}$ & $N_{\tr{PW}}$ &
                $\alpha_{\tr{chg}}$ & $N_{\tr{PW}}$    \\
            \midrule
            All & $0.200000$ & $7^3$ & $0.200000$ & $7^3$ & $0.200000$ & $7^3$ \\
            \bottomrule
        \end{tabular}

        \vspace{1em}

        \begin{tabular}{cccc}
            \toprule
            \multicolumn{4}{c}{GPW} \\
            \midrule
            \multirow{2}*{$N_k$} &
                    C & MgO & LiF   \\
            \cmidrule(lr){2-2} \cmidrule(lr){3-3} \cmidrule(lr){4-4}
            & $N_{\tr{PW}}$ & $N_{\tr{PW}}$ & $N_{\tr{PW}}$ \\
            \midrule
            All & $31^3$ & $83^3$ & $51^3$  \\
            \bottomrule
        \end{tabular}
    \end{table}

    \begin{table}[!h]
        \centering
        \caption{MP2 correlation energies of diamond/cc-pVDZ computed with the HF solution and ERIs calculated using RSGDF and GDF. For $N_k \leq 3^3$, all bands in the cc-pVDZ basis are used, while for $N_k = 4^3$, only the valence occupied bands and the first $10$ conduction bands are correlated to fit the memory limit.}
        \begin{tabular}{cccc}
            \toprule
            $N_k$ & $E^{(2)}_{\tr{RSGDF}}$ / $E_h$ &
                    $E^{(2)}_{\tr{GDF}}$ / $E_h$ &
                    $E^{(2)}_{\tr{RSGDF}} - E^{(2)}_{\tr{GDF}}$ / $\tr{n}E_h$ \\
            \midrule
            $1^3$ & $-0.1702783512$ & $-0.1702783506$ & $-0.58$ \\
            $2^3$ & $-0.2444412960$ & $-0.2444412984$ & $2.40$ \\
            $3^3$ & $-0.2642540235$ & $-0.2642540236$ & $0.13$ \\
            $4^3$ & $-0.2015870651$ & $-0.2015870661$ & $1.03$ \\
            \bottomrule
        \end{tabular}
    \end{table}

    \begin{table}[!h]
        \centering
        \caption{CPU time (unit: second) for computing the Coulomb and exchange matrices in building the Fock matrix per SCF cycle for various methods and systems discussed in the main text. These data are used to make the plots in figure M3.}
        \begin{tabular}{cccccc}
            \toprule
            \multicolumn{6}{c}{Diamond/cc-pVDZ} \\
            \midrule
            $N_k$ & GPW & RSJK & GDF & RSGDF & RSGDF* \\
            \midrule
            $1^3$ & N/A & $1815.25$ & $563.22$ & $14.93$ & $17.17$ \\
            $2^3$ & N/A & $1067.46$ & $855.27$ & $79.43$ & $86.09$ \\
            $3^3$ & N/A & $1452.77$ & $1441.77$ & $205.25$ & $224.41$ \\
            $4^3$ & N/A & $2995.06$ & $2310.46$ & $464.44$ & $481.51$ \\
            $5^3$ & N/A & $4624.48$ & $5063.96$ & $735.42$ & $884.69$ \\
            \bottomrule
        \end{tabular}

        \vspace{1em}

        \begin{tabular}{cccccc}
            \toprule
            \multicolumn{6}{c}{MgO/GTH-DZVP} \\
            \midrule
            $N_k$ & GPW & RSJK & GDF & RSGDF & RSGDF* \\
            \midrule
            $1^3$ & $98.75$ & $5520.58$ & $1320.31$ & $16.28$ & $22.27$ \\
            $2^3$ & $1479.00$ & $4208.34$ & $1743.38$ & $123.39$ & $161.61$ \\
            $3^3$ & $15881.35$ & $4484.17$ & $2459.69$ & $267.78$ & $276.63$ \\
            $4^3$ & $94724.87$ & $6067.74$ & $4534.07$ & $533.97$ & $569.78$ \\
            $5^3$ & $441766.90$ & $6888.72$ & $8343.29$ & $1130.69$ & $1522.41$ \\
            \bottomrule
        \end{tabular}

        \vspace{1em}

        \begin{tabular}{cccccc}
            \toprule
            \multicolumn{6}{c}{LiF/GTH-DZVP} \\
            \midrule
            $N_k$ & GPW & RSJK & GDF & RSGDF & RSGDF* \\
            \midrule
            $1^3$ & $44.23$ & $1929.37$ & $3877.51$ & $29.68$ & $35.42$ \\
            $2^3$ & $203.80$ & $9767.79$ & $6136.51$ & $200.06$ & $221.03$ \\
            $3^3$ & $2442.90$ & $15055.38$ & $8328.24$ & $432.06$ & $516.42$ \\
            $4^3$ & $10362.38$ & $18468.83$ & $11984.96$ & $1073.65$ & $1113.70$ \\
            $5^3$ & $67186.13$ & $20576.87$ & $19416.34$ & $1790.60$ & $1944.22$ \\
            \bottomrule
        \end{tabular}
    \end{table}

    \clearpage


    \section{Treatment of exchange divergence}

    \begin{figure}[!h]
        \centering
        \includegraphics[width=1.0\linewidth]{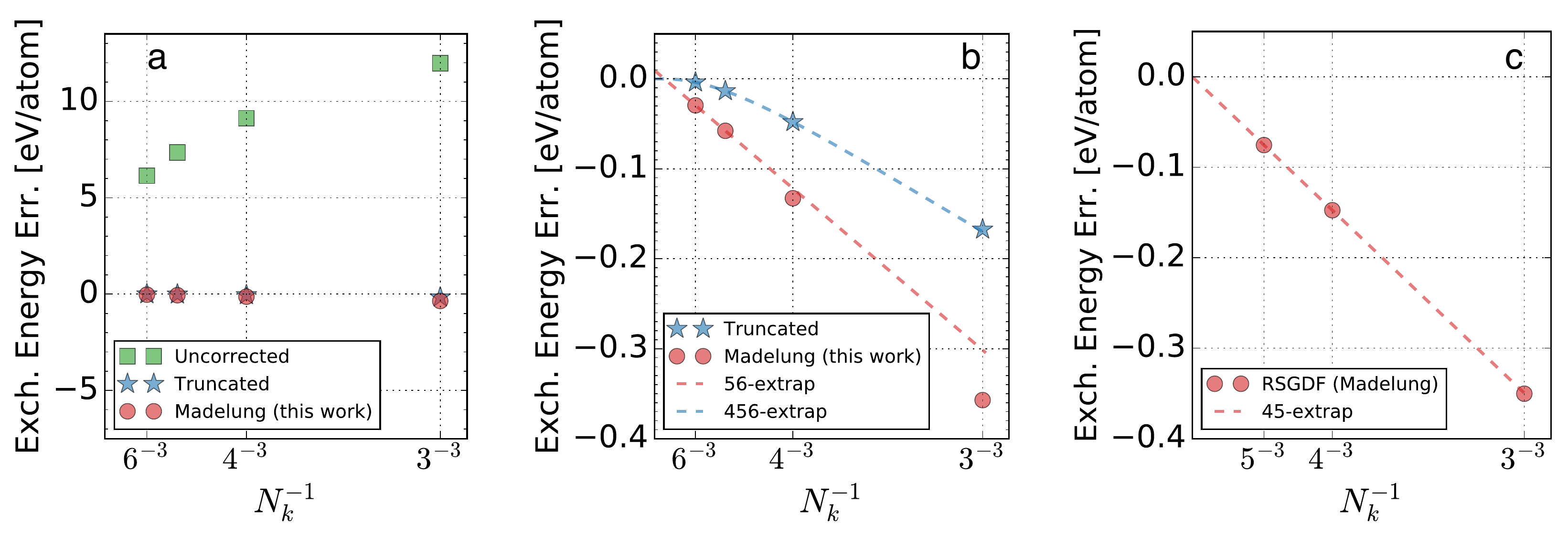}
        \caption{Comparison of different schemes for treating the HF exchange divergence. The HF exchange energies (from converged HF calculations) for diamond calculated using GPW with the GTH-DZVP basis set and the GTH pseudopotential are shown in (a) and (b) and those calculated using RSGDF with the cc-pVDZ basis set are shown in (c). In both (b) and (c), the Madelung results (red circles) from the largest two calculations are used to estimate the TDL result using \cref{eq:exx_Madelung_extrap}, while in (b), the truncated results (blue stars) from the largest three calculations are used for extrapolation using \cref{eq:exx_trunc_extrap}. The $y$-axes in (a) and (b) are shifted by the estimated TDL exchange energy from the truncated scheme, while the $y$-axis in (c) is shifted by the estimated TDL exchange energy from the Madelung scheme.}
        \label{fig:exxtdl}
    \end{figure}

    The $\bm{G} = \bm{0}$ component of the Coulomb potential gives rise to an integrable divergence when evaluating the HF exchange energy in the thermodynamic limit (TDL). However, for any finite $k$-point meshes used to sample the first Brillouin zone, the divergence must be treated explicitly.
    The simplest scheme, which we term \emph{uncorrected}, evaluates the exchange energy as follows
    \begin{equation}    \label{eq:exx_uncorrected}
        E_{\tr{exchange}}^{\tr{uncorr}}(N_k)
            = -\frac{4\pi}{\Omega} \frac{1}{N_k}
            \sum_{i\bm{k}_1}^{\tr{occ}} \sum_{j\bm{k}_2}^{\tr{occ}}
            \sum_{\bm{G}}'
            \frac{
                \rho_{ij}^{\bm{k}_1\bm{k}_2}(\bm{G})
                \rho_{ji}^{\bm{k}_2\bm{k}_1}(-\bm{G})
            }{|\bm{G} + \bm{k}_{12}|^2}
    \end{equation}
    where the primed summation means the $\bm{G} = \bm{0}$ term is omitted when $\bm{k}_{12} = -\bm{k}_1 + \bm{k}_2 = \bm{0}$. The convergence of $E_{\tr{exchange}}^{\tr{uncorr}}(N_k)$ with $N_k$ is very slow, as illustrated in \cref{fig:exxtdl}a for diamond (green squares) using the GTH pseudopotential and the GTH-DZVP basis set.
    GPW is used to handle the ERIs for the results shown in \cref{fig:exxtdl}a and b in order to facilitate the comparison with the \emph{truncated} scheme (\textit{vide infra}) that requires using an auxiliary \emph{plane wave} basis.

    In this work, we follow Ref.~M72 and correct the finite-size error of the exchange energy with the supercell Madelung constant, $E_{\tr{Madelung}}$. This leads to what we called the \emph{Madelung}-corrected exchange energy
    \begin{equation}    \label{eq:exx_Madelung}
        E_{\tr{exchange}}^{\tr{Madelung}}(N_k)
            = E_{\tr{exchange}}^{\tr{uncorr}}(N_k) - \frac{n_{\tr{elec}}}{2} E_{\tr{Madelung}},
    \end{equation}
    where $n_{\tr{elec}}$ is the number of electrons per unit cell. The convergence of $E_{\tr{exchange}}^{\tr{Madelung}}(N_k)$ is $O(N_k^{-1})$ (Ref.~M72), i.e.,
    \begin{equation}    \label{eq:exx_Madelung_extrap}
        E_{\tr{exchange}}^{\tr{Madelung}}(N_k)
            = E_{\tr{exchange}}^{\tr{Madelung}}(\infty)
            + A N_k^{-1}
    \end{equation}
    which is much faster than the uncorrected scheme, as can be seen from \cref{fig:exxtdl}a (red circles).
    Notably, the estimated TDL exchange energy from the Madelung scheme (dashed red line in \cref{fig:exxtdl}b) agrees within $0.01$ eV/atom with that from the so-called \emph{truncated} scheme (dashed blue curve in \cref{fig:exxtdl}b), where the exchange energy is evaluated using a spherically truncated Coulomb potential (Ref.~M75)
    \begin{equation}    \label{eq:exx_trunc}
        E_{\tr{exchange}}^{\tr{trunc}}(N_k)
            = -\frac{4\pi}{\Omega} \frac{1}{N_k}
            \sum_{i\bm{k}_1}^{\tr{occ}} \sum_{j\bm{k}_2}^{\tr{occ}}
            \sum_{\bm{G}}
            \frac{
                \rho_{ij}^{\bm{k}_1\bm{k}_2}(\bm{G})
                \rho_{ji}^{\bm{k}_2\bm{k}_1}(-\bm{G})
            }{|\bm{G} + \bm{k}_{12}|^2} \times
            [1 - \cos(|\bm{G} + \bm{k}_{12}| R_c)]
    \end{equation}
    which leads to an exponential convergence with $N_k$ (Refs.~M75 and M76)
    \begin{equation}    \label{eq:exx_trunc_extrap}
        E_{\tr{exchange}}^{\tr{trunc}}(N_k)
            = E_{\tr{exchange}}^{\tr{trunc}}(\infty)
            + B \exp(- C N_k^{1/3}).
    \end{equation}

    Finally, we show in \cref{fig:exxtdl}c the exchange energy from our RSGDF calculations for diamond in the cc-pVDZ basis set (i.e.,~all-electron). These are the same calculations where the timing data shown in Fig.~M3a are taken. One can see that the exchange energy computed using RSGDF and with the Madelung correction follows the same $1/N_k$ convergence to the TDL as the results shown in \cref{fig:exxtdl}b. We hence conclude that our current treatment of exchange divergence in RSGDF using the Madelung scheme leads to an exchange energy that converges as $O(N_k^{-1})$ to the TDL and can be reliably extrapolated to the TDL using reasonably sized $k$-point meshes.

    \clearpage

    \section{Conditions for prescreening the double lattice summation}

    \begin{figure}[!h]
        \centering
        \includegraphics[width=0.6\linewidth]{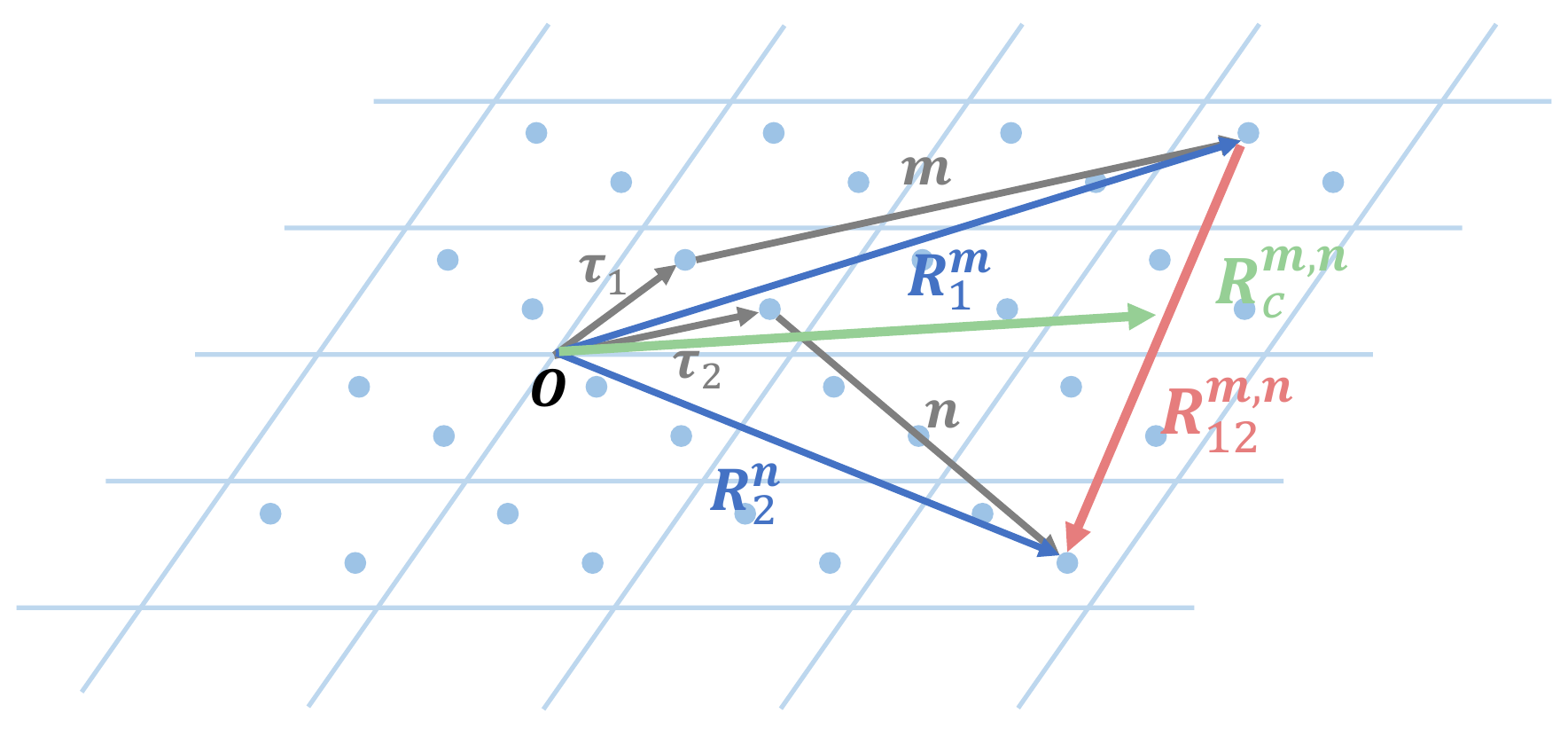}
        \caption{Schematic illustration of the various vectors defined in the text. Note that without loss of generality, the auxiliary orbital $\chi_P$ is assumed to be located at the origin.}
        \label{fig:latvec_illustration}
    \end{figure}

    Here, we derive conditions for prescreening the double lattice summation in Eq.\ (M12), i.e.,\ the short-range (SR) three-center Coulomb integrals. The basic idea is to analyze how each term decays in real space and then derive truncation conditions to achieve a desired precision $\epsilon$ in the computed integrals.

    \subsection{General considerations}

    Eq.\ (M12) can be rewritten as (ignoring the phase factors)
    \begin{equation}    \label{eq:VPmunu}
        V_{P\mu\nu}
            = \sum_{\bm{m},\bm{n}} v_{P\mu\nu}(\bm{R}_1^{\bm{m}}, \bm{R}_2^{\bm{n}};
            \omega)
    \end{equation}
    where
    \begin{equation}    \label{eq:vPmunu_R1R2_gen}
        v_{P\mu\nu}(\bm{R}_1^{\bm{m}}, \bm{R}_2^{\bm{n}}; \omega)
            = \int_{\Omega}\md\bm{r}_1 \int\md\bm{r}_2\,
            \chi_P^{\bm{0}}(\bm{r}_1) w^{\tr{SR}}(r_{12}; \omega)
            \phi_{\mu}^{\bm{R}_1^{\bm{m}}}(\bm{r}_2)
            \phi_{\nu}^{\bm{R}_2^{\bm{n}}}(\bm{r}_2),
    \end{equation}
    $\bm{R}_1^{\bm{m}} = \bm{\tau}_1 + \bm{m}$ and $\bm{R}_2^{\bm{n}} = \bm{\tau}_2 + \bm{n}$, with $\bm{\tau}_1$ and $\bm{\tau}_2$ the position of the two AOs in the reference unit cell (\cref{fig:latvec_illustration}).
    For the simple case of all $1s$-type primitive Gaussians (with exponents $\alpha_{P}$, $\alpha_{\mu}$, and $\alpha_{\nu}$ and coefficients $C_{P}$, $C_{\mu}$, and $C_{\nu}$), the integral in \cref{eq:vPmunu_R1R2_gen} can be easily worked out. The result is
    \begin{equation}    \label{eq:vPmunu_ana}
        v_{P\mu\nu}(\bm{R}_1^{\bm{m}}, \bm{R}_2^{\bm{n}}; \omega)
            = A h (R_{12}^{\bm{m},\bm{n}}) f(R_c^{\bm{m},\bm{n}}; \omega)
    \end{equation}
    where
    \begin{equation}
        A
            = \bigg(\frac{\pi}{4}\bigg)^3
            \frac{C_P C_{\mu} C_{\nu}}{
                \alpha_{P}^{3/2}(\alpha_{\mu} + \alpha_{\nu})^{3/2}
            },
    \end{equation}
    \begin{equation}
        h(R_{12}^{\bm{m},\bm{n}})
            = \me^{-\beta (R_{12}^{\bm{m},\bm{n}})^2}
    \end{equation}
    with $\beta = \frac{\alpha_{\mu}\alpha_{\nu}}
    {\alpha_{\mu} + \alpha_{\nu}}$ and $\bm{R}_{12}^{\bm{m},\bm{n}} = \bm{R}_2^{\bm{m}} - \bm{R}_1^{\bm{n}}$, and
    \begin{equation}
        f(R_c^{\bm{m},\bm{n}}; \omega)
            = \frac{1}{R_c^{\bm{m},\bm{n}}} [
                \tr{erf}(\eta_1 R_c^{\bm{m},\bm{n}}) -
                \tr{erf}(\eta_2 R_c^{\bm{m},\bm{n}})]
    \end{equation}
    with
    \begin{equation}
        \eta_1
            = [\alpha_{P}^{-1} + (\alpha_{\mu}+\alpha_{\nu})^{-1}]^{-1/2},
    \end{equation}
    \begin{equation}
        \eta_2
            = (\eta_1^{-2} + \omega^{-2})^{-1/2}
    \end{equation}
    and
    \begin{equation}
        \bm{R}_c^{\bm{m},\bm{n}}
            = \frac{
                \alpha_{\mu}\bm{R}_1^{\bm{m}} +
                \alpha_{\nu}\bm{R}_2^{\bm{n}}}
            {\alpha_{\mu} + \alpha_{\nu}}.
    \end{equation}
    It can be shown that both $h$ and $f$ are positive-valued, monotonically decreasing functions.

    Define two new lattice vectors,
    \begin{equation}
        \bm{n}'
            = \bm{n} - \bm{m}
    \end{equation}
    \begin{equation}
        \bm{m}'
            = \bm{R}_c^{\bm{m},\bm{m}+\bm{n}'} - \bm{\tau}_{c,\bm{n}'}
    \end{equation}
    where $\bm{\tau}_{c,\bm{n}'}$ lies in the reference unit cell.
    \Cref{eq:VPmunu} can be rewritten as
    \begin{equation}    \label{eq:VPmunu_mnprime}
    \begin{split}
        V_{P\mu\nu}
            &= A \sum_{\bm{n}'} h(R_{12}^{\bm{0},\bm{n'}})
            \sum_{\bm{m}}  f(R_c^{\bm{m},\bm{m}+\bm{n}'}; \omega)   \\
            &= A \sum_{\bm{n}'} h(R_{12}^{\bm{0},\bm{n'}})
            \sum_{\bm{m}'} f(|\bm{m}' + \bm{\tau}_{c,\bm{n}'}|; \omega)
    \end{split}
    \end{equation}
    \Cref{eq:VPmunu_mnprime} changes the double lattice summation in \cref{eq:VPmunu} from over the bare lattice vectors ($\bm{m}$ and $\bm{n}$) to over the center ($\bm{m}'$) and the difference ($\bm{n}'$) vectors.
    From now on, we will drop the prime in the vectors $\bm{m}'$ and $\bm{n}'$.

    \subsection{A bound for $R_c$}

    We first estimate a bound for truncating $\bm{m}$, or equivalently $R_c$. Let $S_h = \sum_{\bm{n}} h(R_{12}^{\bm{0},\bm{n}})$. The error made by truncating the $\bm{m}$-summation is
    \begin{equation}
    \begin{split}
        \delta
            &\approx A S_h \sum_{m > m_{\tr{cut}}} f(m; \omega) \\
            &\approx A S_h \frac{4\pi}{\Omega}
            \int_{R_c^{\tr{cut}}}^{\infty} \md R\, R [
                \tr{erf}(\eta_1 R) - \tr{erf}(\eta_2 R)
            ]   \\
            &\approx A S_h \frac{4\pi}{\Omega} \frac{1}{\sqrt{\pi}}
            \int_{R_c^{\tr{cut}}}^{\infty} \md R\, (
                \eta_2^{-1} \me^{-\eta_2^2 R^2} -
                \eta_1^{-1} \me^{-\eta_1^2 R^2}
            )   \\
            &\leq A S_h \frac{4\pi}{\Omega} \frac{1}{\sqrt{\pi}}
            \frac{1}{R_c^{\tr{cut}}}
            \int_{R_c^{\tr{cut}}}^{\infty} \md R\, R (
                \eta_2^{-1} \me^{-\eta_2^2 R^2} -
                \eta_1^{-1} \me^{-\eta_1^2 R^2}
            )   \\
            &= A S_h \frac{4\pi}{\Omega} \frac{1}{\sqrt{\pi}}
            \frac{1}{R_c^{\tr{cut}}} \frac{1}{2}
            (
                \eta_2^{-3} \me^{-(\eta_2 R_c^{\tr{cut}})^2} -
                \eta_1^{-3} \me^{-(\eta_1 R_c^{\tr{cut}})^2}
            )   \\
            &\approx A S_h \frac{4\pi}{\Omega} \frac{1}{\sqrt{\pi}}
            \frac{1}{2\eta_2^{3}} \frac{1}{R_c^{\tr{cut}}}
            \me^{-(\eta_2 R_c^{\tr{cut}})^2}
    \end{split}
    \end{equation}
    where we have assumed that the final cutoff $R_c^{\tr{cut}}$ is not too small so that
    \begin{enumerate}
        \item in line 1, $f(|\bm{m}+\bm{\tau}_{c,\bm{n}}|; \omega) \approx f(m)$,
        \item in line 2, the summation can be turned into an integral with $(4\pi/\Omega)R^2$ being the density of states,
        \item in line 3, the leading-order asymptotic expansion of the integrand is valid (error $\sim O(R^{-2})$),
        \item in line 6, the $\eta_1$-related term can be dropped.
    \end{enumerate}
    These assumptions could break down if both the AOs and the auxiliary orbitals are very compact so that $R_c^{\tr{cut}}$ is small. However, for virtually all AO bases, even the most compact AOs have non-compact primitive Gaussians that control the long-range behavior and hence the magnitude of $R_c^{\tr{cut}}$. For this reason, the assumptions made above virtually hold for all AO bases. A bound for $R_c$ is hence obtained by requiring
    \begin{equation}    \label{eq:bound_Rc}
    \boxed{
        A S_h \frac{4\pi}{\Omega} \frac{1}{\sqrt{\pi}}
        \frac{1}{2\eta_2^{3}} \frac{1}{R_c^{\tr{cut}}}
        \me^{-(\eta_2 R_c^{\tr{cut}})^2}
            < \epsilon.
    }
    \end{equation}
    Note that for estimating this bound, $S_h$ does not need to be evaluated very accurately. In practice, we fond that performing a small lattice summation
    \begin{equation}    \label{eq:S_h}
        S_h
            \approx \sum_{n < n(\epsilon')} h(R_{12}^{\bm{0},\bm{n}})
    \end{equation}
    suffices, where $n(\epsilon')$ is determined such that $h(R_{12}^{\bm{0},\bm{n}(\epsilon')}) < \epsilon'$. We found that $\epsilon' = 10^{-4}$ already leads to convergent results.

    \subsection{A bound for $R_{12}$}

    We then determine a bound for truncating $\bm{n}$, or equivalently $R_{12}$. First, we note that since the final integral depends on the product of $h$ and $f$, a simple proposal like
    \begin{equation}
        h(R_{12})
            < \epsilon
    \end{equation}
    is not appropriate: it is too tight when $f(R_c;\omega) \ll 1$ (i.e.,\ large $R_c$) and too loose when $f(R_c; \omega) \gg 1$ (i.e.,\ small $R_c$).

    To that end, we determine here a bound for $R_{12}$ that is $R_c$-dependent.
    We first decouple the two summations
    \begin{equation}
        V_{P\mu\nu}
            \leq A \sum_{\bm{m}} f_{\bm{m}}^{\tr{max}}(\omega)
            \sum_{\bm{n}} h(R_{12}^{\bm{0},\bm{n}})
    \end{equation}
    where
    \begin{equation}
        f_{\bm{m}}^{\tr{max}}(\omega)
            = \max_{\bm{\tau} \in \Omega_0}
            f(|\bm{m} + \bm{\tau}|; \omega)
    \end{equation}
    ($\bm{\tau} \in \Omega_{\bm{0}}$ denotes all vectors in the reference unit cell). Now consider the error introduced by truncating the $\bm{n}$-summation for $\bm{m}$ in a finite shell $R \in R + \Delta R$
    \begin{equation}    \label{eq:delta_R_RDeltaR}
        \delta_{R \sim R+\Delta R}
            \leq A N_{\tr{cell}}(R,R+\Delta R) f(R; \omega)
            \sum_{n > n_{\tr{cut}}} h(R_{12}^{\bm{0},\bm{n}})
    \end{equation}
    where $N_{\tr{cell}}(R,R+\Delta R)$ is the number of cells in the range $R \sim R+\Delta R$, and the "$\leq$" comes from the fact that $f$ is a monotonically decreasing function. We found that $\Delta R = 1$ Bohr and
    \begin{equation}    \label{eq:Ncell}
        N_{\tr{cell}}(R,R+\Delta R)
            = \left\{
            \begin{split}
                \frac{4\pi}{\Omega} \times 2 (R+\Delta R)^2 \Delta R,
                    & \quad R > R_0 \\
                N_{\tr{cell}}^{\tr{NF}},
                    & \quad R \leq R_0
            \end{split}
            \right.
    \end{equation}
    works well, where $N_{\tr{cell}}^{\tr{NF}}$ is the number of cells within a sphere with radius $3 \Omega^{1/3}$, and $R_0$ is determined as the crossing point of the two pieces of \cref{eq:Ncell}
    \begin{equation}
        \frac{4\pi}{\Omega} \times 2 (R_0+\Delta R)^2 \Delta R
            = N_{\tr{cell}}^{\tr{NF}}.
    \end{equation}
    The motivation behind \cref{eq:Ncell} is that the continuous approximation of density of states works well for large $R$ but not for small $R$.
    The summation in \cref{eq:delta_R_RDeltaR} can be approximated by an integral
    \begin{equation}    \label{eq:hsum_err}
        \sum_{n > n_{\tr{cut}}} h(R_{12}^{\bm{0},\bm{n}})
            \approx \frac{4\pi}{\Omega}
            \int_{R_{12}^{\tr{cut}}}^{\infty} \md R\, R^2 \me^{-\beta R^2}
            \approx \frac{4\pi}{\Omega}
            \frac{{R}_{12}^{\tr{cut}}}{2 \beta}
            \me^{-\beta (R_{12}^{\tr{cut}})^2}.
    \end{equation}
    Combining \cref{eq:delta_R_RDeltaR,eq:Ncell,eq:hsum_err}, a bound of $R_{12}$ for $R < R_c < R+\Delta R$ is found by
    \begin{equation}    \label{eq:bound_R12_FF}
    \boxed{
        A N_{\tr{cell}}(R,R+\Delta R) f(R;\omega)
        \frac{4\pi}{\Omega}
        \frac{{R}_{12}^{\tr{cut}}}{2 \beta}
        \me^{-\beta (R_{12}^{\tr{cut}})^2}
            < \epsilon.
    }
    \end{equation}
    However, when $R_{12}^{\tr{cut}}$ found by \cref{eq:bound_R12_FF} is too small, the integral in \cref{eq:hsum_err} is not a good approximation to the lattice summation in \cref{eq:delta_R_RDeltaR}. Thus, we use
    \begin{equation}    \label{eq:bound_R12}
    \boxed{
        R_{12}^{\tr{cut}}
            = \max \{\tr{Eq.\ (S23)}, R_{12}^{\tr{cut},\tr{min}}\}
    }
    \end{equation}
    where $R_{12}^{\tr{cut},\tr{min}}$ is determined by
    \begin{equation}    \label{eq:R12cutmin}
        N_{\tr{cell}}(R_{12}^{\tr{cut},\tr{min}},
            R_{12}^{\tr{cut},\tr{min}} + \Delta R)
        \frac{R_{12}^{\tr{cut},\tr{min}}}{2 \beta}
        \me^{-\beta (R_{12}^{\tr{cut},\tr{min}})^2}
            = \sum_{R_{12}^{\bm{0},\bm{n}} > R_{12}^{\tr{cut},\tr{min}}}
            h(R_{12}^{\bm{0},\bm{n}}),
    \end{equation}
    i.e.,\ the minimal $R_{12}$ where the integral in \cref{eq:hsum_err} remains a good approximation to the summation. Similar to $S_h$ in \cref{eq:bound_Rc}, the summation on the right-hand side of \cref{eq:R12cutmin} need not be evaluated very accurately and can be safely truncated using the $n(\epsilon')$ determined for \cref{eq:S_h}.

    \subsection{Summary}

    To summarize, if both the AOs and the auxiliary orbitals are $1s$-type primitive Gaussians,
    \begin{enumerate}
        \item we first use \cref{eq:bound_Rc} to derive a bound for $R_c$, denoted by $R_c^{\tr{cut}}$,
        \item we then make a mesh for $R_c$ with increment $\Delta R$, i.e.,
        \begin{equation}
            0, \Delta R, 2\Delta R, \cdots, R_c^{\tr{cut}},
        \end{equation}
        and use \cref{eq:bound_R12} to determine a series of bounds for $R_{12}$, denoted by $\{R_{12}^{\tr{cut}}(R_c)\}$, each for a window of $R_c$.
    \end{enumerate}
    With these two bounds, a term $v_{P\mu\nu}(\bm{R}_1^{\bm{m}},\bm{R}_2^{\bm{n}}; \omega)$ is not discarded only if
    \begin{equation}
    \begin{split}
        R_c^{\bm{m},\bm{n}}
            &\leq R_c^{\tr{cut}}    \\
        R_{12}^{\bm{m},\bm{n}}
            &\leq R_{12}^{\tr{cut}}(R_c^{\bm{m},\bm{n}})
    \end{split}
    \end{equation}

    The bounds discussed above are determined for every AO shell pair ($\mu\nu$) and auxiliary shell ($P$). For primitive Gaussians with higher angular momentum, we found that treating them as $1s$ works very well (note that the normalization constant needs to be recalculated using the $1s$-orbital normalization condition).
    For a contracted Gaussian orbital, the long-range behavior is determined solely by the most diffuse primitive Gaussian orbital. We can hence use the latter for determining the bounds.

    \clearpage
